\documentclass[aps,pra,superscriptaddress,reprint,longbibliography]{revtex4-1} 
\usepackage{graphicx}
\usepackage{amsmath}
\usepackage[usenames,dvipsnames]{xcolor}
\usepackage[colorlinks=true,linkcolor=blue,urlcolor=blue,citecolor=blue]{hyperref}
\usepackage{braket}
\usepackage{amssymb}
\usepackage{commath}
\usepackage{csquotes}
\usepackage{dcolumn}
\usepackage{bm}
\usepackage{epstopdf}
\usepackage{lipsum}
\usepackage{siunitx}
\usepackage{MnSymbol}
\usepackage{subfigure}
\usepackage{chngcntr}
\usepackage[normalem]{ulem}

\newcommand{\eq}[1]{Eq.~(\ref{#1})}
\newcommand{\fig}[1]{Fig.\thinspace{}\ref{#1}}
\newcommand{\fc}[1]{({#1})}
\newcommand{\figc}[2]{Fig.\thinspace{}\ref{#1}\thinspace{}\fc{#2}}

\newcommand{\rd}{\mathrm{d}}
\newcommand{\bk}{{\bm k}}
\newcommand{\ra}{\rangle}
\newcommand{\bz}{{\bm 0}}
\newcommand{\rra}{\rangle\!\rangle}
\newcommand{\lla}{\langle\!\langle}

\begin{document}
	
\renewcommand*{\theHfigure}{\arabic{figure}} 
	
\title{Interaction dependent heating and atom loss in a periodically driven optical lattice}

\newcommand{\lmu}{Fakult\"{a}t f\"{u}r Physik, Ludwig-Maximilians-Universit\"{a}t M\"{u}nchen, Schellingstr.\ 4, 80799 Munich, Germany}
\newcommand{\mpq}{Max-Planck-Institut f\"{u}r Quantenoptik, Hans-Kopfermann-Str.\ 1, 85748 Garching, Germany}
\newcommand{\pks}{Max-Planck-Institut f\"{u}r Physik komplexer Systeme,
	N\"{o}thnitzer Str. 38,	01387 Dresden, Germany}
\newcommand{\camb}{Cavendish Laboratory, University of Cambridge, J. J. Thomson Avenue, Cambridge CB3 0HE, UK}
\newcommand{\mail}{E-mail: }

\author{Martin Reitter}
\affiliation{\lmu}
\affiliation{\mpq}
\author{Jakob N\"ager}
\affiliation{\lmu}
\affiliation{\mpq}
\author{Karen Wintersperger}
\affiliation{\lmu}
\affiliation{\mpq}
\author{Christoph Str\"ater}
\affiliation{\pks}
\author{Immanuel Bloch}
\affiliation{\lmu}
\affiliation{\mpq}
\author{Andr\'e Eckardt}
\affiliation{\pks}
\author{Ulrich Schneider}
\affiliation{\camb}

\begin{abstract}
Periodic driving of optical lattices has enabled the creation of novel bandstructures not realizable in static lattice systems, such as topological bands for neutral particles. However, especially driven systems of interacting bosonic particles often suffer from strong heating. We have systematically studied heating in an interacting Bose-Einstein condensate in a driven one-dimensional optical lattice. We find interaction-dependent heating rates that depend both on the scattering length and the driving strength and identify the underlying resonant intra- and interband scattering processes. By comparing experimental data and theory, we find that for driving frequencies well above the trap depth, the heating rate is dramatically reduced by the fact that resonantly scattered atoms leave the trap before dissipating their energy into the system. This mechanism of Floquet evaporative cooling offers a powerful strategy to minimize heating in Floquet engineered quantum gases.
\end{abstract}

\pacs{}

\maketitle

\paragraph*{\textbf{Introduction.---}} Floquet engineering, the coherent control of quantum systems by means of time-periodic driving, enables the realization of novel band-structures and many-body phases beyond what is possible in static systems \cite{Zenesini_09, Struck_11, Struck_12, Parker_13, Jotzu_14,Ha_15, Meinert_16, Tai_16, Jotzu_14,Aidelsburger_13b, Aidelsburger_15, Kennedy_15,Struck_13}. It has become an important tool for studies of quantum gases \cite{Eckardt_17}, where it e.g. enables the breaking of time-reversal symmetry and thereby the realization of bands with non-vanishing Chern numbers even for charge-neutral particles \cite{Struck_13, Jotzu_14, Aidelsburger_15, Kennedy_15}. In the form of infrared laser pulses, time-periodic driving can give rise to novel effects in traditional condensed matter systems, such as graphene-like systems \cite{Oka_09, Kitagawa_11, Usaj_14} or high-temperature superconductors \cite{Fausti_11, Matsunaga_14, Mitrano_16}. It also lies at the heart of the recently realized discrete time crystals \cite{Khemani_2016, Keyserlingk_2016, Else_2016, Yao_17, Choi_2017, Zhang_2017}.

Despite those recent accomplishments, successfully combining periodic driving with interactions remains a major experimental challenge, as can be seen already on the level of general thermodynamic considerations: In a driven system energy is not conserved, as the system can absorb or emit energy from or into the drive. Therefore, for any fully ergodic driven system, there can only be one steady state, namely the fully mixed density matrix corresponding to an infinite temperature state \cite{Lazarides_14b, DAlessio_14,Eckardt_08, Poletti_11, Choudhury_14, Choudhury_15, Choudhury_15b, EckardtAnisimovas15, Bilitewski_15a, Bilitewski_15b, Bukov_15, Straeter_16b, Canovi_16}. While this scenario could be avoided by using non-ergodic systems, such as  e.g.\ many-body localized states \cite{Bordia_17}, their use cannot solve the problem in general, as many interesting phases, such as fractional quantum Hall states, are typically ergodic. Therefore, one has to find setups and parameter regimes that allow experimental studies of novel, driven phases on intermediate timescales before the unavoidable heating dominates.

In this work, we experimentally study loss rates of condensed atoms in a driven optical lattice as a function of both driving and interaction strength and can thereby distinguish single-particle from interaction effects. Single-particle
heating occurs via discrete single- or multiphoton interband resonances \cite{Weinberg_2015} that can easily be avoided. This is in contrast to two-particle processes, which in one- or two-dimensional lattices are always resonant, as collisions can convert arbitrary energies into transverse excitations \cite{Choudhury_15b,Bilitewski_15a,Bilitewski_15b}. This is in stark contrast to three-dimensional lattices where these processes can be suppressed  \cite{Eckardt_08,Choudhury_15b,Bilitewski_15a}. In particular, we focus on the two experimentally most relevant driving regimes: For low shaking frequencies $\omega$ much smaller than the resonance frequency to the first excited band but above the bandwidth of the lowest band,  the tunneling matrix element of the lowest band is effectively renormalized by a Bessel function (corresponding to dynamic localization  \cite{Dunlap_86, Sias_08}). At the same time, multiphoton resonances are weak as they require many photons. This regime is typically employed for engineering artificial gauge fields \cite{Jotzu_14,Struck_12,Struck_13}. The second regime lies between the two lowest single-photon single-particle resonances. Here, the dispersion relation can acquire two separate minima that can be exploited to study the formation of symmetry-broken domains \cite{Parker_13, Ha_15}. We find that for large driving frequencies heating is strongly reduced by the fact that scattered particles with energy $\sim\hbar\omega$ typically leave the trap before dissipating the absorbed energy into the system.

\begin{figure}[!t]
	\centering
	\includegraphics[width=.48\textwidth]{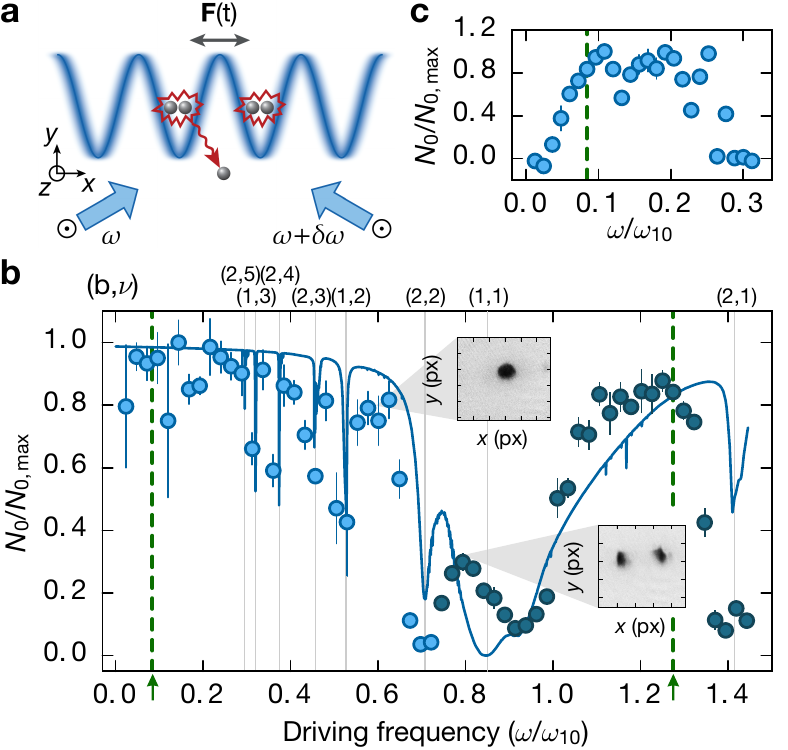}
	\caption{\textbf{Schematic of the experiment and frequency scan.} \textbf{(a)} Two lattice beams with linear out-of-plane polarization intersect at an angle of 120$^\circ$ to form a one dimensional lattice of ``pancakes''. By periodically modulating the frequency of one of the two lattice beams we can shake the lattice, i.e., modulate its position. \textbf{(b)} Normalized atom number after modulating for $50\,\text{ms}$ (for $\omega/\omega_{10}>0.7$) or $100\,\text{ms}$ (for $\omega/\omega_{10}<0.7$) with variable frequency at a driving strength of $\alpha\simeq0.9 $. Error bars indicate the standard error of the mean from four measurements per data point.  The solid blue line shows the theoretically expected single-particle excitations to higher bands. Thin lines mark the resonance positions of multiphoton transitions to higher bands labeled by $(b,\nu)$. Green dashed lines mark the frequencies used in the subsequent study. In the frequency region from roughly $0.7 \, \omega_{10}$ to $1.1 \, \omega_{10}$ we observe a splitting of the BEC due to two degenerate minima in the lowest dressed band, which is included in the theory curve. The insets show raw quasimomentum images of the BEC. \textbf{(c)} Zoom into the regime of small shaking frequencies with $\alpha=2.2$ and $200\, \text{ms}$ shaking duration. }
	\label{sketchandfreqscan}
\end{figure}

\paragraph*{\textbf{Experimental setup.---}} We load an almost pure Bose-Einstein condensate (BEC) of about $4 \times 10^5$ $^{39}$K atoms into the lowest band of a one-dimensional lattice with lattice constant $a=425\, \text{nm}$, which is created by interfering two blue-detuned laser beams with a wavelength of $\lambda=736.8 \, \text{nm}$ at an angle of $\theta$=120$^\circ$, see \figc{sketchandfreqscan}{a}. Then, we shake the lattice position by periodically modulating the frequency of one of the two laser beams. The atoms feel a periodic inertial force in the frame co-moving with the lattice, which is given by $F_x(t)=-(K/a)\cos(\omega t)$, 
where we have introduced the driving amplitude $K$. In order to avoid strong, non-adiabatic excitations to higher bands during the switch-on of the modulation, we continuously ramp the driving amplitude in 10 ms to its desired value. After a variable shaking duration, we determine the heating and losses induced by the drive by measuring the remaining atom number in the BEC. To this end we abruptly stop the drive after an integer number of shaking cycles, immediately followed by bandmapping in the static lattice and 15 ms of time-of-flight (TOF). This TOF is long enough to dilute any thermal background such that we can reliably determine the remaining number of condensed atoms. For all measurements we use a lattice depth of $V_0=11.0(3)\,E_r$, where $E_r$=$h^2 /(8M a^2)\approx h\times 7.1 \, \text{kHz}$ is the effective recoil energy of this lattice with $M$ being the mass of $^{39}$K. In this static lattice, the first interband excitation at zero momentum appears at a frequency of $\omega_{10}=2\pi \times 41.6(5)  \medspace \text{kHz}$.

\paragraph*{\textbf{Frequency scan.---}} 

Single-particle transitions only occur at specific resonances where the shaking frequency $\omega$ fulfills a multiphoton resonance condition $\nu \hbar \omega \simeq \Delta_{b0}(q)$,  with $\nu$ being an integer and $\Delta_{b0}(q)$ denoting the separation of the lowest band to the $b^{th}$ excited band at a given quasimomentum $q$. 
To ensure that we avoid these resonances, we measure the remaining BEC atom number after shaking with variable frequency at a  dimensionless driving strength $\alpha \equiv K/(\hbar\omega)$ and scattering length of $a_s=60\, a_0$, with $a_0$ being Bohr's radius, using a Feshbach resonance at $400\,\text{G}$ \cite{Errico_07}, see \fig{sketchandfreqscan}. The solid blue line shows the result of a numerical single-particle simulation assuming a Gaussian width of the BEC in momentum space of $\Delta q = 0.2 \pi/a$ (for method see \cite{Weinberg_2015, supp}). While the resonances at large frequencies are clearly visible, multiphoton resonances at small driving frequencies are highly suppressed. For our subsequent lifetime measurements we choose $\omega_l=0.084 \, \omega_{10}$ and $\omega_h=1.27 \, \omega_{10}$ (green dashed lines) as low and high frequency, far away from all single-particle resonances.

\begin{figure}[t]
	\centering
	\includegraphics[width=.48\textwidth]{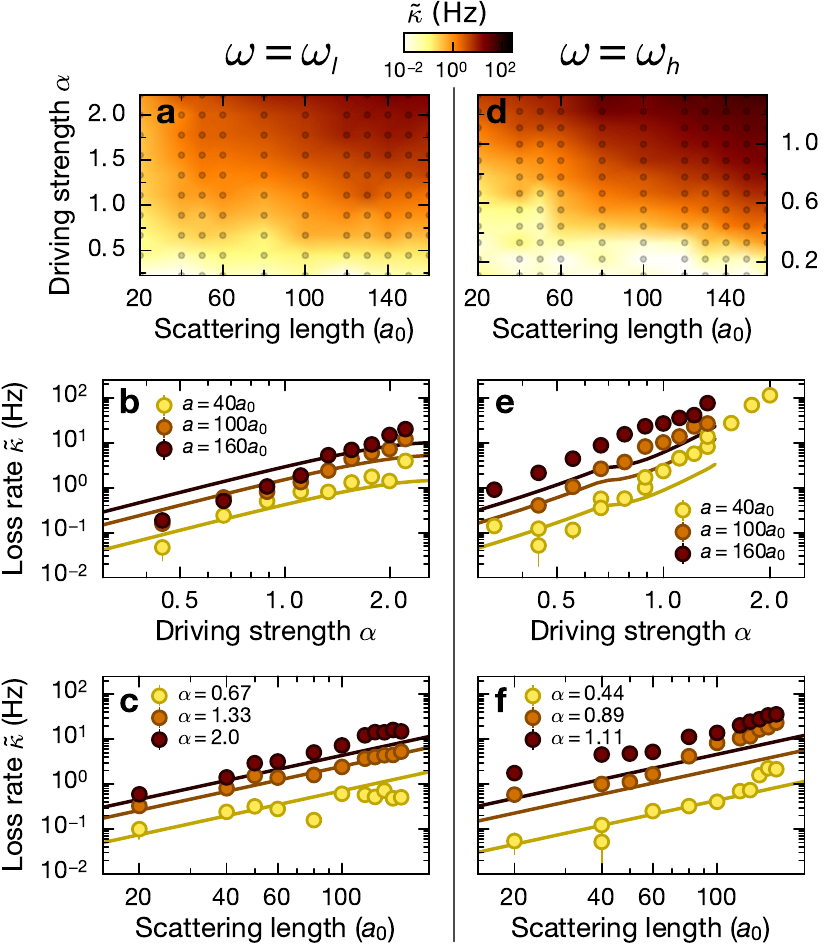}
	\caption{\textbf{Loss rates in the presence of periodic driving.} \textbf{(a,d)} Effective loss rates for different driving amplitudes
		and scattering lengths. Each dot corresponds to a single lifetime measurement. The shaking frequency is (a-c) $\omega=\omega_l$
		and (d-f) $\omega=\omega_h$. \textbf{(b,e)} Crosscuts at fixed scattering lengths.
		The solid lines correspond to the theoretically predicted scattering rates
		and error bars indicate fit uncertainties. \textbf{(c,f)} Corresponding crosscuts at fixed driving strengths. Theory lines in (b,c) assume $f \beta \hbar \omega=10$ to account for the thermalization of scattered atoms (see text).
	}
	\label{Data}
\end{figure}

\paragraph*{\textbf{Experimental loss rates.---}} The total loss rate of condensate atoms in our system is given by summing over background losses in the static system, characterized by a lifetime $\tau$, and heating and losses induced by lattice shaking. We assume that all losses happen on a sufficiently slow timescale such that the system heats up, but stays in global thermal equilibrium, and describe the condensed part using the Thomas-Fermi approximation. We have verified independently that the cloud size indeed shrinks according to the decreasing number of condensed atoms $N_0$ \cite{supp}. Within this approximation, the driving induced loss rate of condensed atoms due to two-particle collisions takes the form $-\kappa N_0^{7/5}$ \cite{supp}. Including the background losses $-N_0/\tau$ of the static system we obtain 

\begin{equation} 
N_0(t)=N_0(0)\frac{e^{-t/\tau}}{\left(1+N_0(0)^{2/5} \kappa \tau \left( 1-e^{-2 t/(5\tau)} \right) \right) ^{5/2}}.
\label{eq:atomdecay}
\end{equation}
We measure $\tau$ independently for each scattering length in the static lattice. In order to form a more intuitive quantity, we define a scaled loss rate $\tilde{\kappa} = \kappa N_0(0)^{2/5}$ such that the initial driving-induced losses scale as $\propto \tilde{\kappa} N_0(0)$. As shown in \fig{Data}, both stronger interactions and larger driving strengths lead to dramatically higher loss rates for the BEC.

\paragraph*{\textbf{Theoretical description.---}} 

\begin{figure}[t]
	\centering
	\includegraphics[width=.48\textwidth]{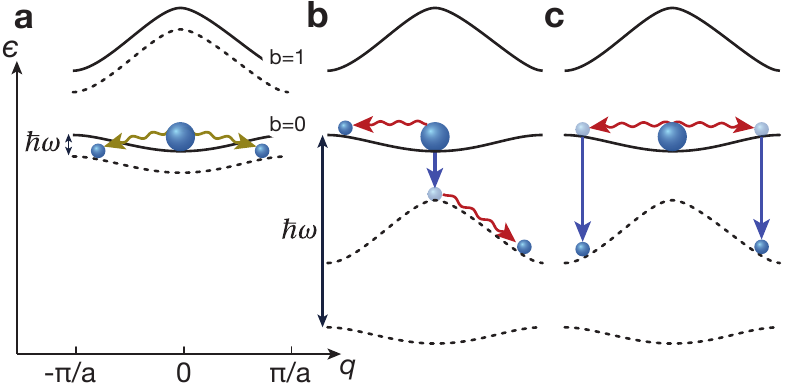}
	\caption{\textbf{Examples of two-particle scattering channels.} The lowest two bands of a schematic lattice dispersion are sketched by solid
		lines, Floquet modes shifted by $-\hbar\omega$ $(m=-1)$ are depicted by dashed lines. The condensate is represented by a large sphere, scattered particles by small spheres. The pair of yellow wiggly arrows in (a) denotes a two-photon scattering process, where the atoms absorb two photons, while the red wiggly lines in (b,c) denote a zero-photon (ordinary) collision between two atoms and blue arrows describe single-photon 
		interband transitions. \textbf{(a)} When the driving frequency is much smaller than the band gap, the dominant loss process are two-photon intraband collisions. \textbf{(b,c)} For driving frequencies larger than the band gap, the leading (subleading) excitation channels combine one (two) single-photon interband transitions with zero-photon collisions. Note that only one photon number $m$ is associated with the whole system and not one per particle as suggested, for simplicity, in the
		diagrams. }
	\label{Heating_processes}
\end{figure}

In order to identify and estimate the dominant heating channels associated 
with two-particle scattering, we start by describing a homogeneous system in the Floquet space of time-periodic states, where an integer Fourier index $m$ describes the change in ``photon'' number relative to a large classical background, i.e., $-m$ counts the number of absorbed photons. In this dressed-atom-like picture, the dynamics is generated by the quasienergy operator $Q$. Within the subspace of a given relative photon number $m$, it acts like $Q_{m,m}=H^{(0)}+m\hbar\omega$, whereas the coupling between 
subspaces $m'$ and $m$ corresponds to an $(m-m')$ photon process which is captured
by $Q_{m',m}=H^{(m'-m)}$. Here $H^{(\nu)}=\frac{1}{T}\int_0^T\!\rd t\,e^{i\nu\omega t}H(t)$ 
denotes the $\nu^{th}$ Fourier component of the time-dependent Hamiltonian $H(t)$. The time-averaged Hamiltonian
$H^{(0)}$ describes a dispersion relation
$\varepsilon_b(k_x)+ E_\perp(k_y,k_z)$, with effective band structure
$\varepsilon_b(k_x)$ and transverse kinetic energy
$E_\perp=\hbar^2(k_y^2+k_z^2)/(2M)$, as well as interactions. 

In \fig{Heating_processes} we sketch the lowest two bands $b=0,1$ for the 
relative photon numbers $m=0,-1$ (solid and dashed lines, respectively),
given by $\varepsilon_b(k_x)+m\hbar\omega$. The diagrams depict 
three examples of relevant scattering channels, where two particles (small spheres) 
are excited out of the condensate into the states $|b,\bk\ra$ and $|b',-\bk\ra$ absorbing $\nu=m-m'$ photons, as indicated by the 
number of particles transferred to the dashed bands ($\nu=1$ in b and
$\nu=2$ in a and c).
The resonance condition for two-particle excitations
can be written as 
\begin{equation}
\varepsilon_b(k_x)+\varepsilon_{b'}(-k_x)-\nu\hbar\omega -2 \varepsilon_0(0)
= - 2E_\perp \le 0. 
\end{equation}
Here, we have separated the transverse kinetic energy
$2E_\perp=E_\perp(k_y,k_z)+E_\perp(-k_y,-k_z)$ created in the scattering process
on one side of the equation. As there is no lattice potential along these directions, the transverse kinetic energy can take arbitrary non-negative values. As a consequence, in \fig{Heating_processes} all states with total energy below the original BEC are accessible.

For the smaller driving frequency $ \omega_l$, the tunneling in the lowest band is modified by a Bessel-function $J_\text{eff} = \mathcal{J}_0(\alpha)J_0$ \cite{supp, Lignier_2007}, where $\mathcal{J}_\nu(\alpha)$ is a Bessel function and $J_0$ the tunneling matrix element of band $b=0$. In this regime, scattering particles to an excited band would require absorbing a large number of photons. Therefore, the dominant heating channel is \textit{intraband} scattering with small $\nu$. Processes with odd $\nu$ are forbidden by symmetry for a condensate with zero momentum, and only acquire small finite values due to the  momentum spread of the condensate. Therefore, two-photon scattering as depicted in \figc{Heating_processes}{a} remains the dominant process in this case. For a $\nu$-photon scattering process the matrix element 
scales like $\sim \hbar^2 a_s n_0 J_0 \mathcal{J}_\nu(\alpha)/(\nu\hbar\omega M)$  \cite{supp}, where $n_0$ is the condensate 
density.

For the larger driving frequency, $\omega_h$, \textit{interband} scattering dominates. In this regime, single-photon single-particle interband coupling is strong (with matrix elements $\sim\alpha E_r$ \cite{supp}) but off resonant. This leads to a 
perturbative admixture of states from the first excited band ($b=1$) with $m=-1$ 
to the lowest band ($b=0$) with $m=0$ and vice versa. As a result
of this coupling, already ordinary zero-photon collisions, which are stronger than $\nu$ photon scattering processes, give rise to excitations by scattering atoms between these dressed bands. Another consequence of this admixture of the highly dispersive first 
excited band to the rather narrow lowest 
band is the formation of a double-well structure within the lowest band for sufficiently large driving strengths $\alpha$
\cite{Parker_13, Ha_15, supp}. As a result, the condensate reforms at the new minima of the dispersion at finite quasimomenta, see insets in \figc{sketchandfreqscan}{b}. We compute the matrix elements for resonant interband excitations using degenerate
perturbation theory. In leading order, we encounter three different single-photon $(\nu=1)$
processes, such as the one depicted in \figc{Heating_processes}{b},
involving a single-particle single-photon interband transition and a zero-photon two-particle scattering event. Their matrix elements scale like 
$\sim \hbar^2a_sn_0 E_r\alpha/(\hbar\omega M)$. The leading correction 
stems from two-photon $(\nu=2)$ processes, an example of which is shown in
\figc{Heating_processes}{c}, giving rise to matrix elements that are a factor
of $\alpha E_r/(\hbar\omega)$ smaller \cite{supp}.

Applying Fermi's golden rule and integrating over the time-dependent Thomas-Fermi profile in a local density approximation, we derive the rates $\Gamma_\nu$ of atoms scattered out of the condensate via $\nu$-photon processes \cite{supp} and find that they are proportional to $(a_s N_0)^{7/5}$. We note that, for the lower driving 
frequency $\omega_l$, the scattered particles will not have enough energy to leave the trap and will  dissipate their entire energy into the system via ordinary (zero-photon) 
collisions. They thereby excite additional atoms out of the condensate, leading to a decay rate for condensed atoms $\dot{N}_0 = -f \beta \hbar\omega \sum_\nu  \nu  \Gamma_\nu /2$ with a numerical factor $f \sim O(1)$ that depends on the details of the system and inverse temperature $\beta$ \cite{supp}. Furthermore, these newly created thermal atoms will continue to absorb energy from the drive. Due to their finite momentum, $\nu=1$ scattering now becomes dominant and thermal atoms will absorb photons at an even faster rate than condensed atoms.
In a truly closed system, this form of heating would scale linearly with the photon energy $\hbar \omega$.
Due to the finite trap depth, however, the system is effectively open and, for the larger driving frequency $\omega_h$, scattered particles typically have sufficient 
energy to quickly leave the trap without dissipating the absorbed energy. In this regime we expect $\dot N_0=-\sum_\nu\Gamma_\nu$.

\begin{figure}[t]
	\centering
	\includegraphics[width=.48\textwidth]{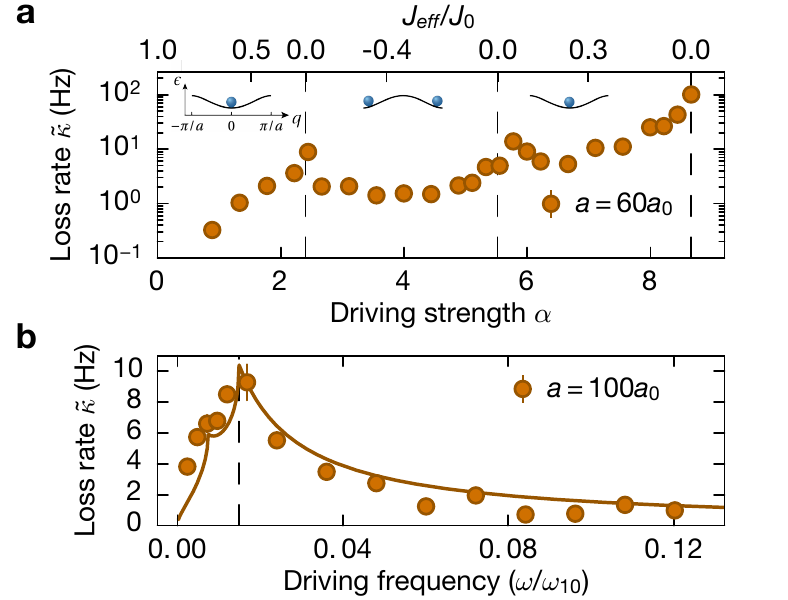}
	\caption{\textbf {Loss rates for large driving strengths and small frequencies} \textbf{(a)} When scanning the driving strength $\alpha$ at a frequency of $\omega_l$, we observe peaks in the effective loss rate whenever the effective tunneling $J_\text{eff}=J_0 \cdot \mathcal{J}_0(\alpha)$ goes to zero (dashed lines). The insets sketch the lowest band for positive and negative tunneling. Error bars indicate fit errors. \textbf{(b)} We observe a peak in the atom number loss rate for a fixed driving strength $\alpha=1.1$ when the driving frequency is close to the bandwidth of the lowest band. The solid line shows the theory scaled by $f \beta \hbar \omega$ with a temperature of $15\, \text{nK}$ (see text). The dashed line indicates the calculated bandwidth of the lattice. }
	\label{amplitude scan}
\end{figure}

\paragraph*{\textbf{Comparison between theory and experiment.---}} Due to the thermalization of the absorbed photon energies described above,
the measured loss rates of condensed
atoms at low driving frequency $\omega_l$ will be
larger than the total scattering rate $\sum_\nu \Gamma_\nu$ (\figc{Data}{a-c}). We observe a factor of $f \beta \hbar \omega \approx 10$, which provides a lower bound for the temperature of the condensate. Assuming for simplicity an ideal homogeneous gas results in a realistic lower bound of $15\,\text{nK}$. 
While typical temperatures of the BEC will likely be higher \cite{supp}, the differences are most likely due to resonant scattering of thermal atoms, which is not included in the theory.
In contrast, for a driving frequency of $\omega_h$, shown in \figc{Data}{d-f}, the loss rate of condensed atoms coincides with the total scattering rate since the absorbed photon energy is carried away with the scattered particles leaving the trap. 
This highlights the advantage of working at larger driving frequencies.

Both \figc{Data}{c} and (f) show that we observe the expected scaling with scattering
length of $\dot{N}_0 \propto a_s^{7/5}$, demonstrating that the
dominant loss mechanisms are indeed interaction driven
and that the Thomas-Fermi local-density approximation
is consistent with our data. 

While the data at large driving frequency $\omega_h$ follows the theory rather well for moderate driving strengths $\alpha$ and scattering lengths, we observe an increasing discrepancy to the expected loss rates for larger scattering lengths, when the mean free path of excited atoms ($\propto 1/a_s^2$) becomes on the order of the size of the BEC. This is most clearly visible when plotting the data vs scattering length, see \figc{Data}{f}, where a discrepancy to the $a_s^{7/5}$ scaling can be observed for scattering lengths larger than $\simeq 100 \, a_0$.  In this regime, excited atoms will undergo additional collisions while leaving the atom cloud, giving rise to an additional loss of condensate atoms similarly to the low frequency case \cite{supp}. We note that two degenerate minima appear in the lowest band for $\alpha>0.7$, giving rise to the small kinks in the expected loss rates
in \figc{Data}{e}. For higher driving strengths we furthermore expect the onset of additional scattering channels with $\nu>2$, which are not included in the theory for $\omega_h$ \cite{supp}.

Finally, we also measured the loss rates for large driving amplitudes and low frequency $\omega_l$, see \figc{amplitude scan}{a}. We can observe clear maxima in the loss rate whenever the effective tunneling matrix element $J_\text{eff}$ is close to zero and attribute them to zero-photon scattering in the effectively flat band. Interestingly, the loss rate decreases again once the sign of the effective tunneling matrix element $J_\text{eff}$ changes. \figc{amplitude scan}{b} shows the loss rate for various frequencies close to the bandwidth. Since there are fewer modes available for frequencies below the bandwidth of the lowest band, a clear decrease in the loss rates can be observed \cite{supp}.

\paragraph*{\textbf{Conclusion and Outlook.---}} We have measured the loss rates of an interacting BEC in a driven one-dimensional optical lattice. We focused on two frequency regimes away from single-particle resonances, which are most relevant for Floquet engineering: Driving frequencies well below the band gap allow for an effective control of tunneling matrix elements and driving frequencies that are blue detuned from the first excited band enable effective bands with two degenerate minima. In both regimes the loss rates approximately scale with the interaction as $a_s^{7/5}$, in agreement with a theoretical description based on a Thomas-Fermi approximation and Fermi's golden rule. We find that for large driving frequencies, scattered particles can leave the trap and carry away
the absorbed energy quanta $\hbar \omega_h$. This mechanism
of continuous Floquet evaporative cooling can act as a powerful
general strategy to reduce heating rates in Floquet
engineered quantum gases. Furthermore, the two-particle scattering processes considered
here rely on exciting transverse motion and might therefore be absent in a three-dimensional lattice. Another intriguing possibility is the use of non-ergodic or many-body localized systems, where the dynamics can be immune to these heating processes.

\paragraph*{\textbf{Acknowledgments.---}} We acknowledge stimulating discussions with Monika Aidelsburger, Marin Bukov, Nigel Cooper, Nathan Goldman and Gaoyong Sun. This work was financially supported by the Deutsche Forschungsgemeinschaft (FOR2414), the European Commission (UQUAM, AQuS) and the Nanosystems Initiative Munich.

\bibliography{shaking_heating}

\cleardoublepage

\appendix

\setcounter{figure}{0}
\setcounter{equation}{0}
\setcounter{page}{1}
\counterwithout{equation}{section}

\renewcommand*{\theHfigure}{\arabic{section}.\arabic{figure}} 

\renewcommand{\thepage}{S\arabic{page}} 
\renewcommand{\thesection}{S.\Roman{section}} 
\renewcommand{\thetable}{S\arabic{table}}  
\renewcommand{\thefigure}{S\arabic{figure}}
\renewcommand{\thesubsection}{\Alph{subsection}}
\renewcommand{\theequation}{S.\arabic{equation}}

\normalem

\def\bea{\begin{eqnarray}}
\def\eea{\end{eqnarray}}
\def\J{\text{J}}

\newcommand{\sgn}{\mathrm{sgn}}
\newcommand{\br}{{\bm r}}
\newcommand{\bq}{{\bm q}}
\newcommand{\bn}{{\bm n}}
\newcommand{\bF}{{\bm F}}
\newcommand{\ri}{\mathrm{i}}
\newcommand{\re}{\mathrm{e}}
\newcommand{\bo}{\hat{b}^{\phantom\dag}}
\newcommand{\ba}{\hat{b}^{\dag}}
\renewcommand{\ao}{\hat{a}^{\phantom\dag}}
\renewcommand{\aa}{\hat{a}^{\dag}}
\newcommand{\no}{\hat{n}}
\newcommand{\Ho}{\hat{H}}
\newcommand{\Uo}{\hat{U}}
\newcommand{\Qo}{\hat{Q}}
\newcommand{\psia}{\hat{\psi}^\dag}
\newcommand{\psio}{\hat{\psi}}
\newcommand{\la}{\langle}
\newcommand{\be}{\begin{equation}}
\newcommand{\ee}{\end{equation}}
\newcommand{\bes}{\begin{eqnarray}}
\newcommand{\ees}{\end{eqnarray}}

\section{Experimental methods}
\setlength{\intextsep}{0.8cm} 
\setlength{\textfloatsep}{0.8cm}

\subsection{\textbf{Preparation of the system}} In order to cool $^{39}$K down to quantum degeneracy we first cool it sympathetically with $^{87}$Rb. In the first cooling stage after laser cooling, we perform microwave evaporation in a plugged quadrupole trap, where both Rb and K are trapped in the $\ket{F=2, m_F=2 }$ hyperfine state. We exploit the different
hyperfine splittings of K and Rb to selectively remove only Rb atoms from the trap by driving the microwave transition between the trapped state $\ket{F=2, m_F=2 }$ and the anti-trapped state $\ket{F=1, m_F=1 }$ of Rb. Due to the small interspecies scattering length between Rb and K microwave evaporation becomes inefficient when reaching the low $\mu$K regime \cite{Sarlo_07}. We therefore transfer both species into a crossed beam dipole trap where we can tune the interspecies interaction with a Feshbach resonance \cite{Ferlaino_06}. After loading both species into the dipole trap we transfer them into their absolute ground states $\ket{F=1, m_F=1 }$ and set the interspecies scattering length to $a_{K,Rb}\approx100 \, a_0$. By lowering the intensity of the dipole beams, the selective evaporation of Rb continues due to the larger gravitational sag and hence weaker vertical trapping potential for Rb. After loosing all Rb atoms we continue the evaporation with K alone. In this last stage of the evaporation we set the scattering length to $a_{K,K}\approx 150 \,a_0$ and cool the atoms down to an almost pure BEC. After the evaporation we ramp the Feshbach field to its desired value in $50\, \text{ms}$. Finally, we load the atoms into the lattice by linearly ramping up the lattice beams in $100 \, \text{ms}$ to their final lattice depth of $11 \, E_r$. By performing Bloch oscillations with subsequent bandmapping we determine the momentum spread of the condensate to be approximately $\Delta q=0.2 \pi/a$.
\\

\subsection{\textbf{Atom number calibration}}
\begin{figure}[htbp]
	\centering
	\includegraphics[width=.4\textwidth]{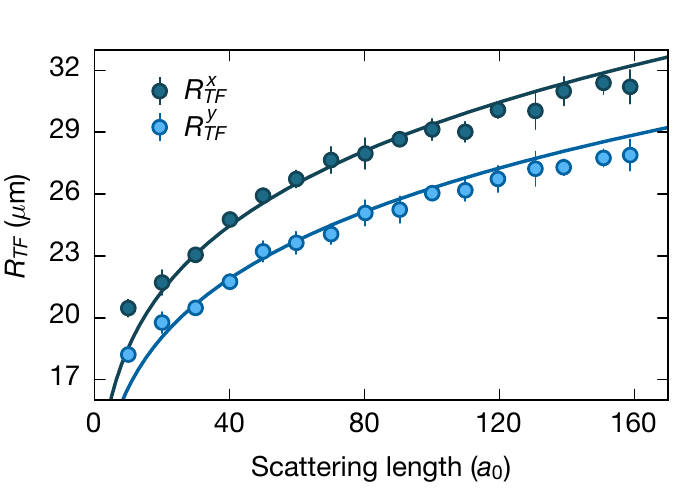}
	\caption{\textbf{Atom number calibration} Measured insitu Thomas-Fermi radii along the $x-$ and $y-$axis. By fitting \eq{eq:atomnumberfit} to the data we can calibrate the atom number of our system. Error bars indicate the standard deviation of the mean from six individual measurements per data point.}
	\label{appendix:atnumcal}
\end{figure}

As the atom number directly enters the fit function used to determine the decay rates it is crucial to calibrate the atom number as well as possible. Since the initial BEC in the dipole trap can be described accurately within the Thomas-Fermi approximation, we can employ it to calibrate the atom number. According to the Thomas-Fermi model in a harmonic trap the condensate radius $R_{TF}^{i}$ with $i={x,y,z}$ is given by
\begin{equation}
R_{TF}^i=\Bigl(\frac{15 N_0 a_s \hbar^2 \omega_x \omega_y \omega_z }{M^2 \omega_i^5}\Bigr)^{1/5}.
\label{eq:atomnumberfit}
\end{equation}
As we can measure the harmonic trapping frequencies and the insitu Thomas-Fermi radius of our system very precise and can, in addition, tune the scattering length using a Feshbach resonance, we are able to calibrate the atom number by fitting \eq{eq:atomnumberfit} to our data. From this measurement we extract the scaling factor between the measured optical densities in our time-of-flight (TOF) images and the real atom number. \fig{appendix:atnumcal} shows the fit to the measured insitu radii of the BEC. Each point is an average over 6 individual shots leading to an atom number of $4.0(2)\times 10^5$ atoms.
\\

\subsection{\textbf{Fitfunction}\label{Lossmechanisms}}
\begin{figure}[htbp]
	\centering
	\includegraphics[width=.4\textwidth]{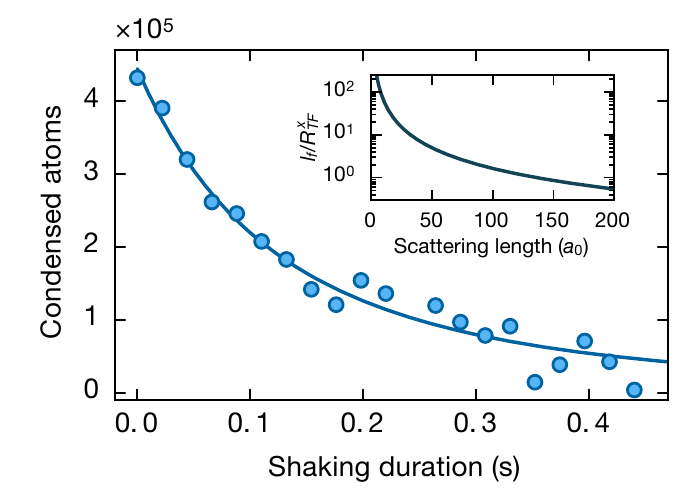}
	\caption{\textbf{Atom number decay versus shaking duration} An example of a decay rate measurement. The driving strength was $\alpha = 1$ at a shaking frequency of $\omega = \omega_h $ and a scattering length of $a_s=80 \, a_0$. The inset shows the ratio between mean free path and condensate radius  $l_f/R_{TF}^x$. For large scattering lengths the mean free path becomes on the order of the size of the BEC resulting in a finite probability of collateral scattering events. }
	\label{exampledecay}
\end{figure}

Starting from the Thomas-Fermi ansatz, the atom number loss from the BEC due to the shaking is described by the differential equation $\dot{N}_0=-\kappa N_0^{7/5}$ (for derivation see \ref{sec:trap}). This differential equation has the solution: 

\begin{equation} 
N_0(t)=N_0(0)\frac{1}{(1+\frac{2}{5}\kappa t N_0(0)^{2/5})^{5/2}}.
 \label{eq:shakingdecay}
\end{equation}
However, this decay function only describes atom number losses due to the shaking of the lattice. In addition we also have technical heating and losses due to collisions with the hot background gas that need to be taken into account. These losses are described by the differential equation $\dot{N}_0=-\kappa_{bg} N$. As a good approximation we can assume that these two loss channels are independent of each other leading to $\dot{N}_0=-\kappa_{bg} N_0 -\kappa N_0^{7/5}$. The total atom number loss is then given by

\begin{equation} 
N_0(t)=N_0(0)\frac{e^{-t/\tau}}{\left(1+N_0(0)^{2/5} \kappa \tau \left( 1-e^{-2 t/(5\tau)} \right) \right) ^{5/2}},
\label{eq:atomdecayappendix}
\end{equation}
where $\tau=1/\kappa_{bg}$ is the measured lifetime in the static lattice. Since $\tau$ becomes shorter for larger scattering lengths we measure the lifetime in the static lattice for each scattering length. An exemplary measurement from which we infer the decay rate $\kappa$ is shown in \fig{exampledecay}.\\

\subsection{\textbf{Validity of Thomas-Fermi approach}}
\begin{figure}[htbp]
	\centering
	\includegraphics[width=.47\textwidth]{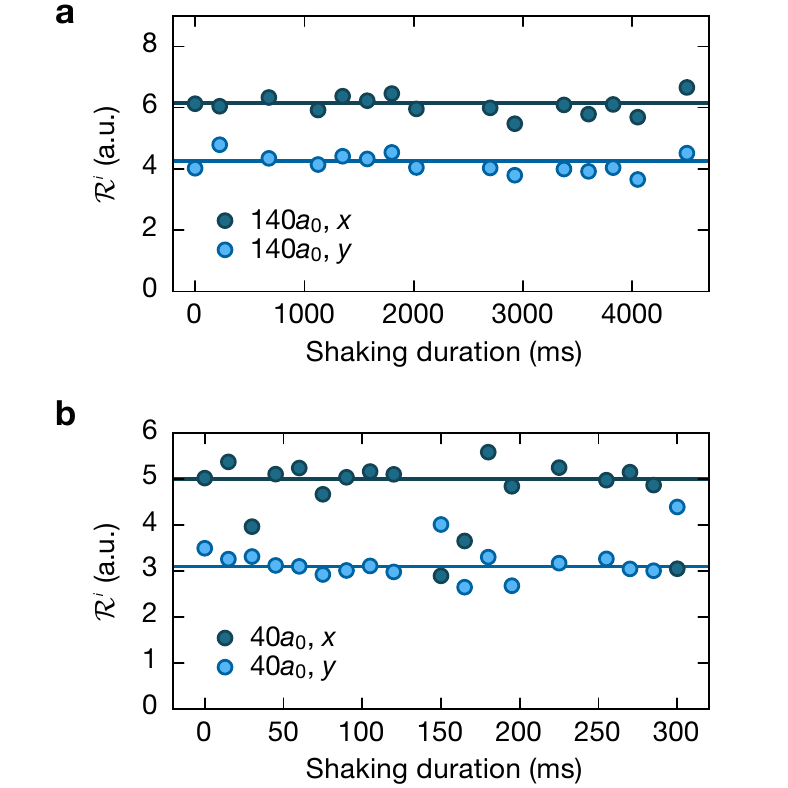}
	\caption{\textbf{Check on thermal equilibrium} \textbf{(a)} The ratio $R_{TF}^i/N_0(t)^{1/5}$ at $\omega=\omega_l$, $\alpha=0.44$ and $a_s=140\,a_0$ is plotted versus the shaking duration.  \textbf{(b)} $\mathcal{R}^i$ for $\omega=\omega_h$, $\alpha=1.3$ and $a_s=40\,a_0$. Solid lines are a guide to the eye.}
	\label{fig:thermaleq}
\end{figure}

The fit function \eq{eq:atomdecayappendix}, from which we infer the loss rates $\kappa$ of condensed atoms, assumes that the system stays in a global thermal equilibrium while the lattice is being shaken. This implies that the BEC is being described by a Thomas-Fermi model during most parts of the shaking process. According to \eq{eq:atomnumberfit} this means that the ratio $\mathcal{R}^i=R_{TF}^i/N_0(t)^{1/5}$, with $i={x,y}$, stays constant. In order to verify this, we measure the insitu Thomas-Fermi radius for various shaking durations and directly afterwards repeat the same measurement but this time determine the remaining atom number in the BEC from TOF images. \fig{fig:thermaleq} shows exemplary measurements of $\mathcal{R}^i$ for two different shaking parameters. In \figc{fig:thermaleq}{a} the shaking frequency is set to $\omega=\omega_l$ with a driving strength $\alpha=0.44$ and a scattering length of $a_s=140\,a_0$. For these shaking parameters the lifetime is fairly long and the system is expected to stay in a global thermal equilibrium. This is confirmed by the measurement as $\mathcal{R}^i$ stays constant during the whole loss process. In \figc{fig:thermaleq}{b} we show a more extreme example where the scattering length is set to $a_s=40\,a_0$ and thus thermalization processes due to collisions take more time. We shake the system at a frequency of $\omega=\omega_h$ with a driving strength of $\alpha=1.3$ which leads to a large loss rate of condensed atoms. Even in this more extreme case $\mathcal{R}^i$ stays constant during the shaking process. 

\subsection{\textbf{Harmonic trapping frequencies}} We determine the harmonic trapping frequencies in the dipole trap by giving the atoms a kick and observing the resulting oscillating motion in the trap. We measure trapping frequencies of  $\omega_x=2\pi \times 24.2(1)\, \text{Hz}$,  $\omega_y=2\pi \times 27.6(4)\, \text{Hz}$ and  $\omega_z=2\pi \times 204(3)\, \text{Hz}$. To be able to compare the measured decay rates with theory (section \ref{sec:theory_description}), we however require the trapping frequencies in the presence of the lattice. A direct measurement is hampered by the fact that the lattice damps dipole oscillations rather quickly. We can therefore only measure the trapping frequency $\omega_z$ along the vertical direction, giving $\omega_z=2\pi \times 186(3)\, \text{Hz}$. 
The trapping frequency along the $y$-direction is calculated by modeling the anticonfinement due to lattice beams for the $11E_r$ lattice. Combining this with the measured trapping frequency in the dipole trap results in $\omega_y=2\pi \times 25.2 \, \text{Hz} $. The ratio of the trapping frequencies $\omega_x/\omega_y$ is equal to the ratio of the Thomas-Fermi radii $R_{TF}^x/R_{TF}^y$. From the measured in situ cloud shape the harmonic trapping frequency along the lattice direction can be estimated to be $\omega_x=2\pi \times 18.6 \, \text{Hz} $. From the beam parameters of the dipole and lattice beams we can estimate the trap depth along the vertical direction as $V_\text{ver} \approx h\times  20\, \text{kHz}$ and  $V_\text{hor} \approx h\times 18 \,\text{kHz}$  along the horizontal direction.
\\
\subsection{Temperature estimation}

For the smaller driving frequency $\omega_l$, scattered atoms will typically not have enough energy to leave the trap and hence will redistribute their acquired energy over the system via rapid zero-photon collisions. This redistribution leads to an additional loss of condensate atoms, such that the total rate becomes $\dot{N}_0 = -f \beta \hbar\omega \sum_\nu  \nu  \Gamma_\nu /2$. We empirically found $f \beta \hbar \omega_l \approx 10$. For an ideal homogeneous three dimensional Bose gas the numerical factor $f$ is given by $f \approx 0.78$, see \ref{sec:multiplicationeffect}. In the spirit of the local density approximation we refrain from using the factor $f$ of a trapped ideal Bose gas. Ignoring the enhanced scattering rate of thermal atoms, the above empirical factor provides a lower bound of the temperature of $T\approx15 \, \text{nK}$. When taking bandmapped TOF pictures of the condensate we cannot detect any thermal background, from which we can conservatively infer that the condensate fraction will be above $90\%$. For the homogeneous ideal Bose gas the condensate fraction is described by $N_0/N=1-(T/T_c)^{3/2}$ \cite{Bagnato_87}, where $T_c$ is the critical temperature. Inserting a condensate fraction of $>90\%$ in this formula yields an upper bound for the  temperature  of  $T<60\,\text{nK}$, which corresponds to $f \beta \hbar \omega>2$, compatible with the measured factor of $\approx 10$. Note that the measured loss of condensate atoms tends to overestimate $f\beta\hbar\omega$, since in the course of time also thermal (non-condensed) atoms will start to absorb photons in resonant collisions.  
Due to their finite momentum, single-photon processes are not suppressed anymore. These collisions are even faster than the two-photon processes giving the main contribution to the resonant scattering of two condensate atoms, leading to an even faster energy absorption from the drive.

\subsection{\textbf{Mean free path}}
For the high shaking frequency $\omega_h$ scattered particles have enough energy to leave the trap and hence the predicted theoretical scattering rate should coincide with the measured loss rate. This is, however, only true if scattered particles can leave the trap before additional collisions with condensed atoms occur. In order to estimate the probability of such additional collisions, which increase the measured loss rates, we estimate the mean free path $l_f=1/\bar{n}\sigma$ of excited atoms. Here $\sigma=8\pi a_s^2$ is the scattering cross section and $\bar{n}=N_0/V$ denotes the mean density of the BEC. Since we can describe the density distribution of the BEC by a Thomas-Fermi model, the condensate volume is given by $V=4\pi/3 R_{TF}^x R_{TF}^y R_{TF}^z$. The inset of \fig{exampledecay} shows the ratio of the mean free path to the Thomas-Fermi radius along the $x-$direction. For large scattering lengths the mean free path becomes on the order of the size of the BEC, leading to a finite probability of scattering events between excited and condensed atoms. Due to the high energy of the excited atoms, potentially a large number of condensed atoms can be excited, leading to a significantly higher loss rate.

\subsection{\textbf{Detection and analysis}} After shaking the atoms for a variable duration in the lattice, we suddenly stop the shaking after an integer number of periods and perform bandmapping in the static lattice by ramping down the lattice beams in $100\,\mu$s. To count the remaining number of atoms in the BEC, we perform absorption imaging of the cloud after $15\, \text{ms}$ TOF. Especially for the lower shaking frequency $\omega_l$ , where the scattered atoms do not leave the trap but dissipate the absorbed photon energy into the system, a large thermal background is formed. After $15\, \text{ms}$ TOF this thermal background is already strongly diluted and hence can be assumed to be almost homogeneous in the vicinity of the condensate. In order to only count condensed atoms, we  choose a second region of interest (ROI) that
is close to but separated from the condensate peak and subtract the mean density within this background ROI from every pixel of the main ROI. 

\section{Theoretical estimation of condensate depletion}\label{sec:theory_description}

\subsection{The system}
We describe the system of bosonic atoms in a shaken one-dimensional lattice 
in the reference frame comoving with the lattice. The Hamiltonian reads
\be
\Ho(t) = \int\!\rd \br \bigg[\psia(\br)h(\br,t)\psio(\br) 
            + \frac{g}{2}\psia(\br)\psia(\br)\psio(\br)\psio(\br)\bigg].
\ee
with single-particle Hamiltonian
\be
h(\br,t) = -\frac{\hbar^2}{2M}\nabla^2+V_L(\br)-\bF(t)\cdot\br +V_\text{trap}(\br).
\ee
Here $M$ denotes the atomic mass, $V_L(\br) = V_0\sin^2(k_Lx)$ a one-dimensional
(1D) optical lattice potential, $\bF(t)=-\frac{K}{a}\cos(\omega t) {\bm e}_x$ the
shaking-induced homogeneous inertial force, and
$V_\text{trap}(\br) =\frac{1}{2}m(\omega_x^2 x^2+\omega_y^2y^2+\omega_z^2 z^2)$
the trapping potential. The field operators $\psio(\br)$ and $\psia(\br)$ describe 
the annihilation and creation of a boson at position $\br$, respectively. The 
interactions are described by the coupling constant $g=4\pi\hbar^2 a_s/m$, where
the $s$-wave scattering length $a_s$ can be tuned using a Feshbach resonance. 
In the following we will consider parameter values corresponding to the 
experiment: $V_0/E_r=11$ as well as two driving frequencies, the lower one
$\hbar\omega/E_r=0.5$ and the larger one $\hbar\omega/E_r=7.5$. Both the 
dimensionless driving amplitude $\alpha=K/(\hbar\omega)$ and the dimensionless 
scattering length $a_s/a$ are varied.

For our analysis we will first neglect the impact of the trap and consider a 
translational invariant system with linear extent $L$ in all three 
directions and periodic boundary conditions. The role of the spatial confinement 
will then be estimated in a subsequent step using a local-density approximation based on 
the Thomas-Fermi wave function for the condensate. The single-particle problem 
separates with respect to the three spatial directions. In the transverse 
directions $\br_\perp = (y,z)$, the stationary states $|\bk_\perp\ra$ are 
characterized by sharp momentum wave numbers
$\bk_\perp =(k_y,k_z)$, which take discrete values $k_i=2\pi \nu_i/M$ with 
integers $\nu_i$. Their wave functions and energies read
\be 
\la\br_\perp|\bk_\perp\ra = \frac{1}{L}\exp(i\bk_\perp\cdot\br_\perp)
\quad\text{and}\quad
E_\perp(\bk_\perp) = \frac{\hbar^2 \bk_\perp^2}{2M}.
\ee
Along the lattice direction the undriven system is described by the single-particle 
$ h_x^{(0)} = -(\hbar^2/2M)\partial_x^2 + (V_0/2)\cos(2\pi x/a)$.
In the absence of driving the eigenstates  $|b k_x\ra$ are Bloch waves
characterized by both a band index $b=0,1,2,\ldots$ and a quasimomentum wave
number $k_x$, taking discrete values $k_x=2\pi \nu_x a/L$, with integers
$\nu_x$, so that $-\pi<a k_x\le \pi$. Their energies $E_b(k_x)$ define the band
structure and their wave functions read
\be
\la x|bk_x\ra = u_{bk_x}(x)e^{ik_xx}, \quad  u_{bk_x}(x)=u_{bk_x}(x+a).
\ee

The Bloch states $|bk_x\ra$ are superposition of momentum eigenstates
$|\beta k_x\ra$ with momenta $k_x + \beta 2\pi/a$,
which differ from the quasimomentum by an integer $\beta$ times of reciprocal
lattice constants $2\pi/a$,
$|bk_x\ra = \sum_\beta u_{b\beta}(k_x) |\beta k_x\ra $.
In terms of the momentum eigenstates, which are plane waves,
$\la x|\beta k_x\ra=\exp\big(ix (k_x +\beta \frac{2\pi}{a}\big)/\sqrt{L}$,
the undriven single-particle Hamiltonian takes the form
$\la \beta'k_x'|h_x^{(0)}|\beta k_x\ra =\delta_{k_x'k_x}
h^{(0)}_{\beta'\beta}(k_x)$,
with
\be
h^{(0)}_{\beta'\beta}(k_x) = 
\frac{\hbar^2}{2M}\Big(k_x+\frac{2\pi}{a}\beta\Big) \delta_{\beta'\beta} 
+ \frac{V_0}{4}\big(\delta_{\beta',\beta+1}+\beta_{\beta',\beta-1}\big).
\ee

The Bloch waves can also be expressed in terms of Wannier states $|b\ell\ra$,
$|bk_x\ra = \sqrt{\frac{a}{L}}\sum_{\ell=1}^{L/a} e^{ik_x \ell a}|b\ell\ra$,
whose wave functions $\la x |b\ell\ra=w_b(x-\ell a)$ are real, exponentially
localized at the lattice minima $x=\ell a$, and obey $w_b(-x)=(-)^b w_b(x)$.
The lowest band, and to lesser extent also the first excited band, can be
approximately described in a tight-binding description, where the kinetics is
dominated by tunneling between (Wannier states of) neighboring lattice sites. 
Denoting the corresponding tunneling matrix elements of the lowest two bands by
$-J_b$ the dispersion relation takes the form
\be\label{eq:Eb}
E_b(k_x) \simeq \varepsilon_b -2J_b\cos(ak_x),
\ee
where $\varepsilon_b$ denotes the band-center energy, corresponding to the orbital
energy of the localized Wannier state. One finds that $J_b$ is positive (negative)
for $b$ being even (odd). The kinetics of the higher excited bands,
with $b\ge 3$ and to lesser extent with $b=2$, resembles rather that of free
particles with sharp momentum, except for quasimomenta close to 
$0$ or $\pi/a$, where Bragg reflection hybridizes states of opposite momenta. 
These bands are not captured by Eq.~(\ref{eq:Eb}). 

In terms of the Bloch states, the interactions read
\be\label{eq:Hint_bk}
\Ho_\text{int} = \frac{g}{2L^3}\sum_{\{b\bk\}} \zeta_{\{b\bk\}}
\aa_{b_4\bk_4}\aa_{b_3\bk_3}\ao_{b_2\bk_2}\ao_{b_1\bk_1}
\ee
with dimensionless coefficients
$\zeta_{\{b\bk\}} \equiv \zeta_{b_4\bk_{4},b_3\bk_3,b_2\bk_2,b_1\bk_1}$, 
\bes\label{eq:zeta_bk}
\zeta_{\{b\bk\}}
&&= \sum_{\{\beta\}} 
u_{b_4\beta_4}^*(k_{x4}) u_{b_3\beta_3}^*(k_{x3}) 
u_{b_2\beta_2}(k_{x2}) u_{b_1\beta_1}(k_{x1}) 
\nonumber\\&&\times\,
\delta_{k_{x1}+k_{x2}+(\beta_1+\beta_2)\frac{2\pi}{a},
	k_{x3}+k_{x4}+(\beta_3+\beta_4)\frac{2\pi}{a}}
\nonumber\\&&\times
\delta_{\bk_{\perp 1}+\bk_{\perp 2},\bk_{\perp 3}+\bk_{\perp 4}}.
\ees
Below, we will consider scattering matrix elements for scattering of two 
particles having zero quasimomentum, $\bk={\bm q}$, into states with quasimomenta 
${\bm q}\pm\bk$. These shall be denoted by
$\zeta_{b_4b_3b_2b_1}(\bk,\bq) = \zeta_{b_4\bq+\bk,b_3\bq-\bk,b_2\bq,b_1\bq}$
as well as by $\zeta_{b_4b_3b_2b_1}(\bk)\equiv\zeta_{b_4b_3b_2b_1}(\bk,\bz)$. 
If we can neglect the interactions between particles occupying Wannier states
at different lattice sites, which is a good approximation for the lowest band(s),
the coefficients $\zeta_{\{b\bk\}}$ become independent of quasimomentum,
$\zeta_{\{b\bk\}} \simeq \zeta_{\{b\}}$, and read
$\zeta_{b_4b_3b_2b_1} 
= a\int\!\rd x\, w_{b_4}(x)w_{b_3}(x)w_{b_2}(x)w_{b_1}(x)
$.
Thanks to on-site parity conservation, within this approximation
$\zeta_{b_4b_3b_2b_1}$ vanishes when $b_1+b_2-b_3-b_4$ is odd. For convenience, 
we also define $\zeta\equiv \zeta_{0000}$.

\subsection{The driven single-particle problem}

For non-zero driving strength $K$, the system possesses generalized stationary
states called Floquet states. The scalar potential $-xF(t)$ that incorporates the 
homogeneous driving force apparently breaks the lattice symmetry. 
However, this symmetry can be restored by applying a gauge
transformation, so that the Floquet states are Bloch waves characterized by sharp 
quasimomenta (Floquet-Bloch states). 
We will consider different gauge transformations that restore the lattice
symmetry, depending on the driving frequency.

\subsubsection{Driving frequencies well below the band gap}\label{sec:sp_low}

In the regime, 
where the driving amplitude $K$ and frequency $\hbar\omega$ are 
smaller than the separation $\Delta_{10}$ of the lowest two bands
the most dominant effect of the forcing on the states of the lowest band will be
a periodic translation in \emph{quasimomentum} by
\be\label{eq:q(t)}
q(t)=\frac{1}{\hbar}\int_0^t\rd t' F(t') = -\frac{\alpha}{a} \sin(\omega t)
\ee
that conserves the band index $b$.

Such a translation is described by the time-periodic unitary operator
\be\label{eq:U1}
U_1(t) = \sum_{b k_x} |b k_x+q(t)\ra\la b k_x| 
=  \exp\bigg(i a q(t)  \sum_{b\ell} \ell  |b\ell\ra\la b\ell| \bigg).,
\ee
It defines a gauge transformation giving rise to the transformed Hamiltonian,
$h'_x = U_1^\dag h_x U_1-i\hbar U_1^\dag\dot U_1$.
It reads
\bes\label{eq:h'}
h'_x(t) = \sum_{k_x}\bigg[\sum_b E_b\big(k_x+q(t)\big)|bk_x\ra\la b k_x|
\nonumber\\
+\, K\cos(\omega t) \sum_{b'b} \eta_{b'b}(k_x,t)
|b'k_x\ra\la b k_x|    \bigg].
\ees
where we have introduced the dimensionless interband coupling parameter
\be
\eta_{b'b}(k_x,t) = \sum_{\Delta \ell} e^{-i\Delta \ell[k_x+q(t)]}
\eta_{b'b}^{(\Delta \ell)}
\ee
with $\eta_{b'b}^{(\Delta \ell)}= \int\!\rd x\, 
w_{b'}(x-\Delta \ell a) \frac{x}{a} w_b(x)$.
For the coupling between the two lowest bands, for which the Wannier functions
are well localized, we might approximate the sum by the $\Delta\ell=0$ term,
$\eta_{10}(k_x,t)=\eta_{01}(k_x,t)\simeq \eta_{10}^{(0)}$.

In a first approximation, we can neglect the coupling to excited bands and so
that the states of the lowest band form a band of Floquet-Bloch states whose
quasienergies $\varepsilon_0(k_x)$ are given by the time-averaged dispersion
\be\label{eq:E0'}
\varepsilon_0(k_x) = \frac{1}{T}\int_0^T\!\rd t\, E_0\big(k_x+q(t)\big)
\simeq \epsilon_0 -2J_0^\text{eff} \cos(ak_x) 
\ee
[as well as copies of that band shifted by integer multiplies of the energy
$\hbar\omega$, $\varepsilon_{0m}(k_x) = \varepsilon_0(k_x)  + m\hbar\omega$].
On the right side of Eq.~(\ref{eq:E0'}), we have introduced the effective 
tunneling matrix element $J_0^\text{eff}=J_0\mathcal{J}_0(\alpha)$.
This result is obtained by employing Eq.~(\ref{eq:Eb}) and the identity 
$ \exp\big(ia\sin(b)\big)=\sum_k \mathcal{J}_k(a)\exp(ikb)$, with Bessel function 
of the first kind $\mathcal{J}_k(a)$. Transforming back to the original frame of 
reference (co-moving with the lattice) by employing $U_1(t)$, the so-approximated 
Floquet-Bloch state is described by an accelerated Bloch state $|b k_x+q(t)\ra$.

\begin{figure}[t]
	\includegraphics[width=0.49\linewidth]{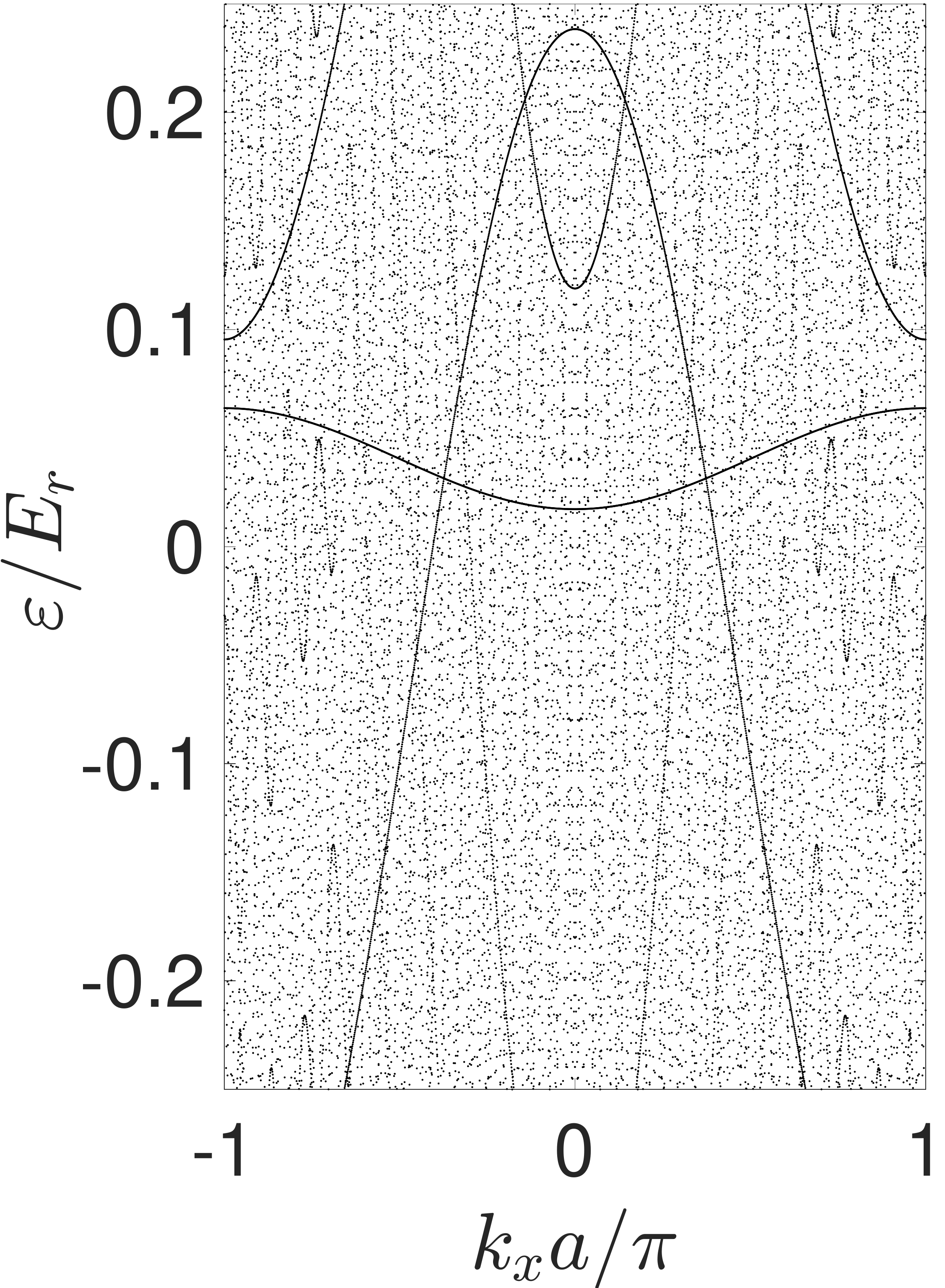}
	\includegraphics[width=0.49\linewidth]{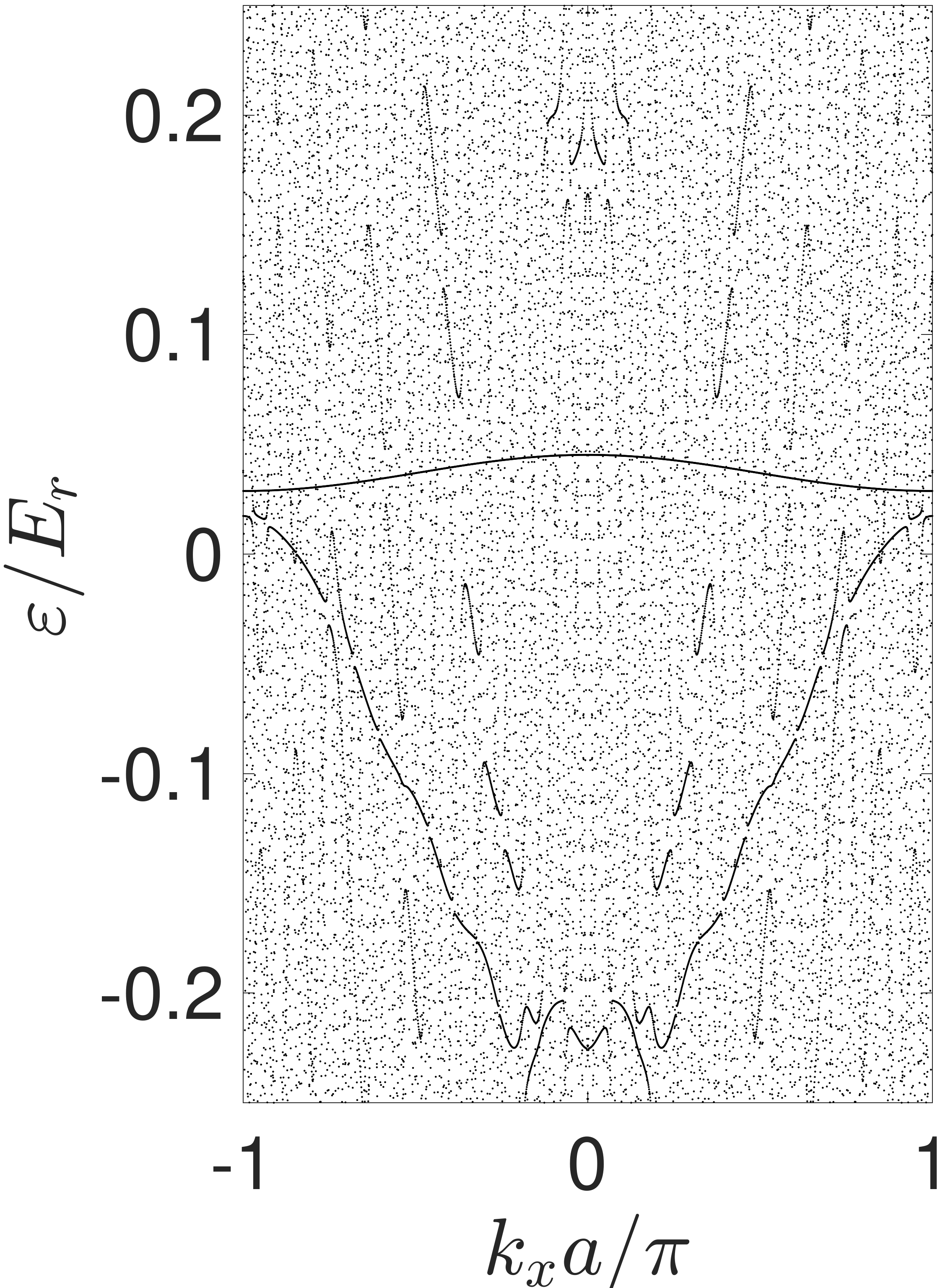}
	\centering
	\caption{\label{fig:QBands_low} 
		One Brillouin zone of the quasienergy band structure of the driven cosine
		lattice with $V_0/E_r=11$, $\hbar\omega/E_r=0.5$, and $\alpha = 1$ (left panel)
		as well as $\alpha=3$ (right panel).} 
\end{figure}

In Fig.~\ref{fig:QBands_low} we plot one Brillouin-zone (i.e.\ an interval of 
width $\hbar\omega$) of the quasienergy spectrum of the driven
lattice with $\hbar\omega/E_r=0.5$ for two different driving strengths,
$\alpha=1$ (left panel) and $\alpha =3$ (right panel). For these values the
tunneling matrix element is effectively modified by a factor of
$\mathcal{J}_0(1)\approx0.77$ and $\mathcal{J}_0(1)\approx-0.26$, respectively.
At the bottom of both Figure panels, we can clearly recognize a quasienergy
band originating from the lowest band of the undriven system. In the right panel
one can, moreover, clearly see the effective inversion of this band, as it is
expected for $J_0^\text{eff}<0$. This confirms the reasoning underlying the
approximation (\ref{eq:E0'}).

\subsubsection{Driving frequencies above the band gap}\label{sec:sp_high}

\begin{figure}[t]
	\includegraphics[width=0.49\linewidth]{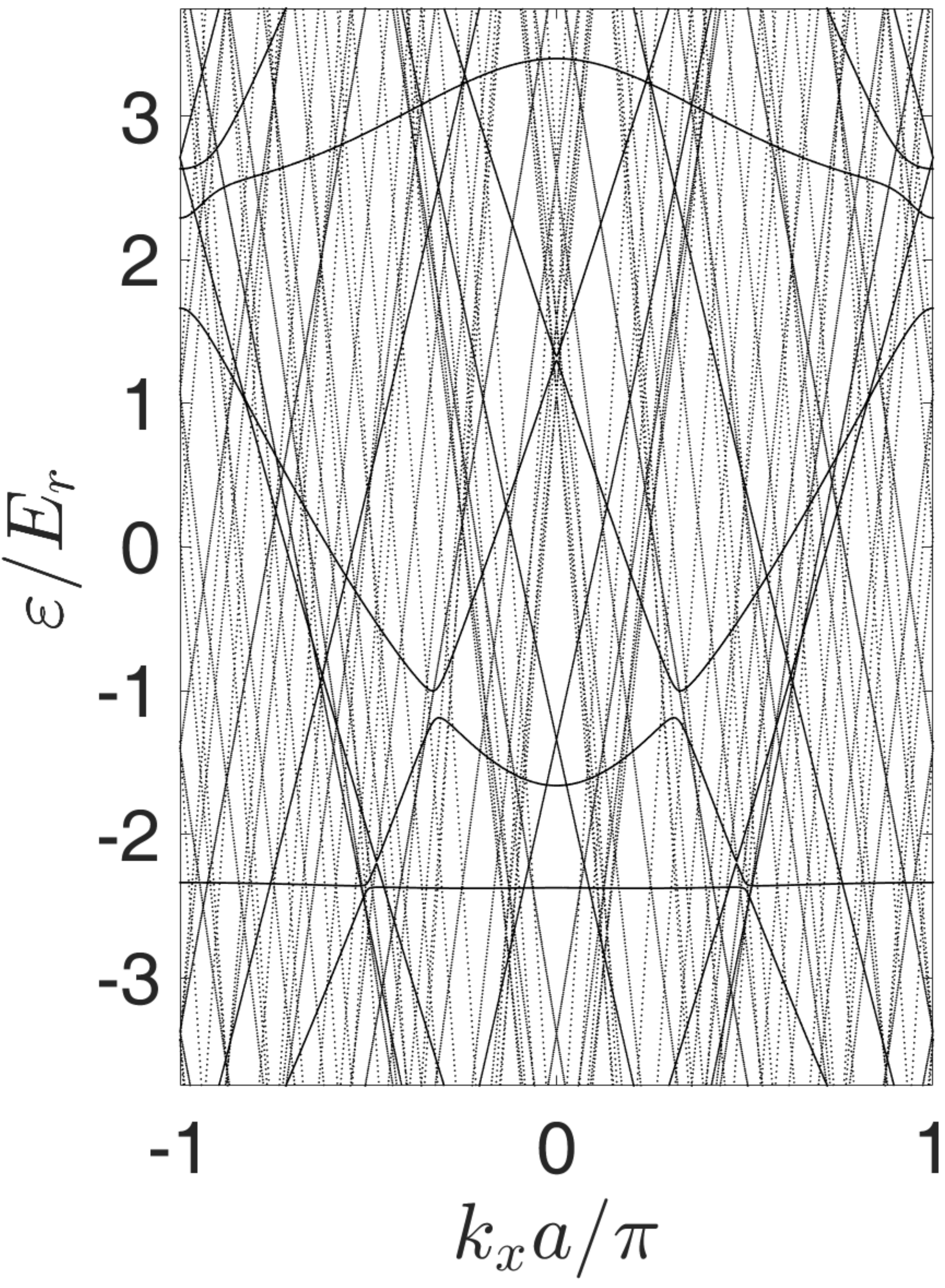}
	\includegraphics[width=0.49\linewidth]{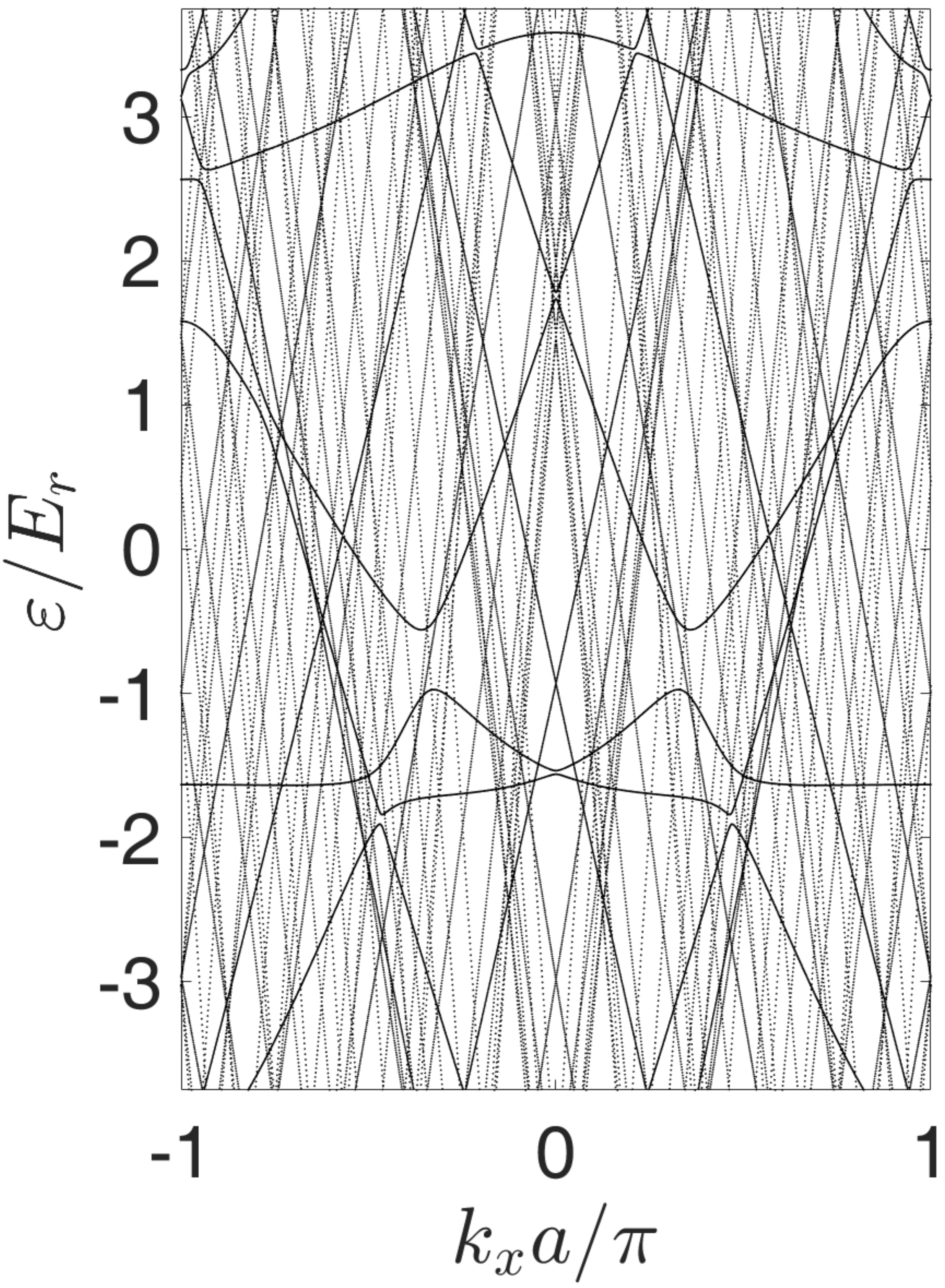}
	\centering
	\caption{\label{fig:QBands_high} 
		One Brillouin zone of the quasienergy band structure of the driven cosine
		lattice with $V_0/E_r=11$, $\hbar\omega/E_r=7.5$, and $\alpha = 1$ (left panel)
		as well as $\alpha=3$ (right panel).} 
\end{figure}

For driving frequencies and amplitudes that are large compared to the first band gap
the driving will strongly mix the undriven bands. The system rather follows 
momentum eigenstates than quasimomentum eigenstates (i.e.\ Bloch states). In 
this regime it is convenient to perform a gauge transformation with the unitary 
operator 
\be\label{eq:U2}
U_2(t) = \exp\big(i q(t)x\big)
\ee
describing a translation by $q(t)$ in \emph{momentum}. The transformed
Hamiltonian $h''_x = U_2^\dag h_x U_2-i\hbar U_2^\dag\dot U_2$ reads
\bes\label{eq:h''}
h''_x(t) = \frac{\hbar^2}{2m}\big[-i\partial_x + q(t)\big]^2 + V_0\sin^2(k_L x)
\nonumber\\
= h^{(0)}_x + \frac{\hbar^2}{2m} \big[2q(t)(-i\partial_x) + q^2(t)\big].
\ees

Neglecting the coupling terms involving $q(t)$ in the Hamiltonian
(\ref{eq:h''}), the eigenstates of the system are simply given by the undriven
Bloch states, so that we expect quasienergy bands described by the dispersion 
relation
\be\label{eq:Eb''}
\varepsilon_{b}(k_x) \simeq E_b(k_x)
\ee
(as well as copies of them shifted by integer multiplies of the energy
$\hbar\omega$,
$\varepsilon_{bm}(k_x) = \varepsilon_b(k_x)  + m\hbar\omega$].
This approximation is not only valid for the lowest band(s), but for all bands. 
In fact it accurately describes the physics of highly excited bands with 
energies well above the lattice depth $V_0$, which are separated from each other
by tiny band gaps only. Transforming back to the original frame of reference
(co-moving with the lattice) by employing $U_2(t)$, the Floquet-Bloch state is 
described by momentum shifted
Bloch states $e^{ixq(t)}|b k_x\ra$.

In Fig.~\ref{fig:QBands_high} we plot one Brillouin-zone of the quasienergy
spectrum of the driven lattice with $\hbar\omega/E_r=7.5$ for two different
driving strengths, $\alpha=1$ and $\alpha =3$. In the left panel, we can clearly
identify copies of the lowest, as well as of the first and second excited bands.
Even higher excited bands are also present, but harder to identify due to
their steeper dispersion. In the right panel we can see that these bands can
still be identified for the stronger driving amplitude (even though the lowest 
band now undergoes large avoided crossings with excited bands).

\subsubsection{Multi-photon interband excitations}

The interband-coupling terms appearing in the Hamiltonians (\ref{eq:h'}) and
(\ref{eq:h''}) can induce transitions between different Bloch bands. This can
happen even when the driving frequency $\hbar\omega$ is small compared to the 
band gap, $\hbar\omega<\Delta_{10}$. Such single-particle processes conserve 
quasimomentum. Assuming a particle in the $k_x=0$ state of the lowest band, the 
excitation to the first excited band is, thus, connected to the resonance 
condition
\be
\nu\hbar\omega \approx \varepsilon_1(0)-\varepsilon_0(0)     
\ee 
for an $\nu$-photon transition, where $\varepsilon_{0\nu}(0)
=\varepsilon_0(0)+\nu\hbar\omega$ and $\varepsilon_{10}(0)=\varepsilon_1(0)$
are (nearly) degenerate. Such multi-photon interband 
heating processes have been investigated theoretically and  experimentally
in Refs.~\cite{Weinberg_2015, Straeter_16b}). For driving amplitudes below 
a threshold value, they are suppressed exponentially for large $n$. These
resonances are clearly visible also in the data presented in \figc{sketchandfreqscan}{b} of
the main text. The theoretical curves in this plot were obtained by integrating 
the single-particle time evolution starting from undriven Bloch states of the 
lowest band. The plotted line corresponds to the minimal probability for 
remaining in this state encountered during the time evolution, averaged over a 
group of momenta $k$ representing the measured momentum distribution.

\subsection{Heating rates from two-particle scattering}
Driving induced heating occurs, when the system absorbs energy quanta
$\hbar\omega$. One source of heating are the single-particle processes mentioned
at the end of the previous section, which occur at driving frequencies close to 
sharp resonance conditions. We find them to be suppressed in two regimes: for
sufficiently small driving frequencies $\hbar\omega\ll\Delta_{10}$, i.e.\ for
large photon numbers $n$, as well as for driving frequencies
$\hbar\omega>\Delta_{10}$ lying in the second band gap. In these two regimes, 
the main heating channels are given by different types of interaction-induced
resonant scattering processes out of the condensate. In the following we will 
estimate the corresponding loss rates of condensed atoms. 
Resonant scattering between Floquet-Bloch states, has recently also been described 
in Refs.~\cite{Choudhury_14, Choudhury_15, Bilitewski_15a, Bilitewski_15b, 
GenskeRosch15}. (For a discussion of heating processes in the strongly coupled Mott 
insulator regime, see Refs.~\cite{Eckardt_08, EckardtAnisimovas15}).

\subsubsection{General form of the scattering rates}
In the following we will assume that the system of $N$ particles is prepared in
a low-entropy state with a Bose condensate of $N_0\approx N$ atoms in the
single-particle ground state. When the driving is switched on, experimental data 
shows that the condensate is transferred to  the Floquet Bloch state 
$|b=0,\bk=\bq\ra$ at the minimum of that Floquet-Bloch band originating from the 
lowest undriven band. Starting from this state, we compute the matrix elements
$C^{(\nu)}_{bb'}(\bk,\bq)$ for the (dominant) $\nu$-photon scattering processes 
into the target state where two condensate atoms have been transferred into the 
Floquet-Bloch states $|b,\bq+\bk\ra$ and $|b,\bq-\bk]\ra$ with
$\bk=(k_x,\bk_\perp)$. For the given contact interactions the matrix element
$C^{(\nu)}_{bb'}(\bk,\bq)$ will not depend on the transverse momentum
$\bk_\perp$ and it will be proportional to $g N_0/L^3$, so that
\be
C^{(\nu)}_{bb'}(\bk,\bq) \equiv C^{(\nu)}_{bb'}(k_x,q)
\equiv \frac{N_0g}{L^3} c_{b'b}^{(\nu)}(k_x,q),
\ee
with dimensionless intensive factor $c_{b'b}^{(\nu)}(k_x)$.

The accessible states are defined via the condition 
$
\varepsilon_b(q+k_x)+\varepsilon_{b'}(q-k_x) +2E_\perp(\bk_\perp)
- 2\varepsilon_0(q) - \nu\hbar\omega = 0$.
Thanks to the continuum of transverse modes $\bk_\perp$ the transverse energy
$2E_\perp(\bk_\perp)$ can take any non-negative value, so that it translates to
\be\label{eq:condition}
\varepsilon_b(q+k_x)+\varepsilon_{b'}(q-k_x) - 2\varepsilon_0(q)
 -\nu\hbar\omega < 0. 
\ee
For convenience, we define that $c^{(\nu)}_{bb'}(k_x,q)=0$ whenever the resonance 
condition (\ref{eq:condition}) is not fulfilled. For the sake of a light notation, 
below we will suppress the arguments $\bq$ and $q$ in the following, unless we are 
explicitly considering the case $\bq\ne\bz$, 
which becomes relevant when discussing the case of larger driving frequencies. 

The rate for $\nu$-photon scattering out of the condensate can then be estimated 
by employing the golden rule
\be
\Gamma_\nu = 2\frac{1}{2}  
\sum_{bb'} \sum_{k_x} \frac{2\pi}{\hbar}|C^{(\nu)}_{b'b}(k_x)|^2\rho_\perp
= 32N_0 (n_{2D}a_s^2) \frac{E_r}{\hbar} \gamma_\nu
\ee
Here the prefactor of 2 accounts for the fact that atoms are scattered pairwise,
the factor of $1/2$ takes into account that $(b,k_x)\leftrightarrow (b',-k_x)$ 
does not lead to a new target state, and $\rho_\perp=ML^2/(2\pi\hbar^2)$ 
denotes the density of states with respect to the two-dimensional space given 
by the transverse directions. In the second step we introduced the condensate 
density $n_{2D}=a N_0/L^3$ in the 2D plane defined by every lattice minimum
and the dimensionless scattering parameter 
\be\label{eq:gamma}
\gamma_\nu  =\sum_{b'b}\frac{a}{2\pi}\int_{-\frac{\pi}{a}}^{\frac{\pi}{a}}                       \! \rd k_x |c_{b'b}^{(\nu)}(k_x)|^2,
\ee
by replacing $\sum_{k_x}\to \frac{L}{2\pi}\int\! \rd k_x$.

\subsubsection{Taking into account the trap}\label{sec:trap}
In order to take into account the effect of the trapping potential, we will
employ two approximations, the Thomas-Fermi approximation and the local-density 
approximation. Within the \emph{Thomas-Fermi approximation} one computes the 
condensate wave function by neglecting both the depletion and the kinetic energy. 
In our lattice system, the starting point is a description of the system in terms 
of the Wannier states of the lowest band. In this approximation the order 
parameter of the condensate takes the form $\psi_\ell(y,z)=\sqrt{n_\ell(y,z)}$, 
where $n_\ell(y,z)$ is the 2D condensate density in the $\ell$th 
lattice minimum, i.e.\ in the Wannier states $|0\ell\ra$. The potential energy of 
the system is given by 
\be
E = \sum_\ell\int\!\rd y\rd z 
\bigg[ \frac{\tilde{g}}{2 a} n_\ell(y,z) 
+V_\text{trap}(\ell a,y,z)-\mu\bigg]n_\ell(y,z),
\ee
where $\mu$ is the chemical potential and $\tilde{g}=g\zeta$ the coupling 
constant. It is convenient to 
approximate the site index $\ell$ by the continuous position
$x=a\ell$, so that $\sum_\ell\frac{1}{a}\to\int\!\rd x$, and to define the
three-dimensional condensate density
$n(\br)=n_{x/a}(y,z)/a$. In this description the energy becomes minimal for the density profile
\be
n(\br) = \left\{\begin{array}{ll}
	\frac{\mu-V_\text{trap}(\br)}{\tilde{g}} & \text{ where } V_\text{trap}(\br)<\mu
	\\
	0 & \text{ elsewhere}.\\
\end{array}\right.
\ee 
Here the chemical potential $\mu$ is determined by the total number of particles
in the condensate, from $N_0 = \int\!\rd \br \,n(\br)$ we obtain
\be
\mu =   \bigg[\frac{15\pi}{8} \zeta \frac{a_s}{a} 
\bigg(\frac{\hbar\bar{\omega}}{E_r}\bigg)^3N_0\bigg]^{2/5}E_r,
\ee
where $\bar{\omega}=(\omega_x\omega_y\omega_z)^{1/3}$.

Within a \emph{local-density approximation} the total scattering rate is now
given by integrating over the scattering rate per volume,
$\Gamma_\nu/L^3$, 
\bes\label{eq:GammaLDA}
\Gamma_\nu^\text{LDA} &=& 32 \gamma_\nu a a_s^2 \frac{E_r}{\hbar} \int\!\rd \br\, n^2(\br)
\\\nonumber
&=&  \frac{128}{105}\zeta^{-3/5}
\left(\frac{15\pi}{8}\frac{a_s}{a}N_0\right)^{7/5} 
\left(\frac{\hbar\bar{\omega}}{E_r}\right)^{6/5}
\frac{E_r}{\hbar}\gamma_\nu.
\ees

\subsubsection{Multiplication effect via thermalization}\label{sec:multiplicationeffect}

At least as long as the driving frequency is small compared to the trap depth,
the scattered condensate atoms will dissipate the absorbed photon energy
$\nu\hbar\omega$ energy into the system by undergoing rapid ordinary zero-photon 
scattering processes with other atoms, leading to a multiplication effect. We can 
estimate this effect by assuming that the system equilibrates immediately.
For an ideal Bose gas with density of states 
$ 
g(E)\equiv b E^\gamma
$ 
and inverse temperature $\beta$, the depletion of the condensate,  $N'=N-N_0$,
is given by
\be
N'=  \int_0^\infty \! \rd E\, g(E) \frac{1}{e^{\beta E}-1}
= \frac{b \Gamma(\gamma+1)\zeta(\gamma+1)}{\beta^{\gamma+1}}
\ee
where $\Gamma(x)$ and $\zeta(x)$ denote the Gamma and the Zeta function, 
respectively. The energy is given by
\be
E = \int_0^\infty \! \rd E\, g(E) \frac{E}{e^{\beta E}-1}
= \frac{b \Gamma(\gamma+2)\zeta(\gamma+2)}{\beta^{\gamma+2}}.
\ee
Thus,
\be
N'  = \underbrace{\frac{\Gamma(\gamma+1)}{\Gamma(\gamma+2)}}_{(\gamma+1)^{-1}}
\frac{\zeta(\gamma+1)}{\zeta(\gamma+2)}\beta E
\propto E^{\frac{\gamma+1}{\gamma+2}}
\ee
and, with that,
\be
\frac{\rd N' }{\rd E} 
= \frac{1}{\gamma+2}\frac{\zeta(\gamma+1)}{\zeta(\gamma+2)} \beta \equiv
f \beta .
\ee
For the homogeneous three dimensional Bose gas, one has $\gamma = 1/2$, giving
$f\approx 0.78$ and for the three-dimensional harmonic oscillator one finds
$\gamma=2$ and $f\approx 0.28$. For the interacting system in the trap, the 
depletion sees a potential given by a combination of the trap and a repulsive 
central bump given by the condensate, so that $\gamma=2$ should not be a very
good approximation for low-energy states. Nevertheless one can expect a factor
$f$ of the order of one. 
Noting that a pair of atoms that is scattered out of the condensate via an
$\nu$-photon process receives an energy of $\nu\hbar\omega$ to be dissipated in 
the system, the depletion increases by about $(\nu/2)f \beta \hbar\omega$ 
for each scattered atom. Thus, we can estimate the loss of condensate atoms to
be given by 
\be\label{eq:Ndotlow}
\dot N_0 =- \sum_{\nu>0} \frac{\nu f}{2} \beta\hbar\omega\Gamma_\nu
\ee
The factor $\beta\hbar\omega$ can be explained intuitively, by noting that the
mean energy of an excited atom is $\sim\beta^{-1}$.

In our system, we find that for the higher driving frequency the scattered 
condensate atoms have gained sufficient energy to leave the trap and to carry away the photon energy $\nu\hbar\omega$. In this regime we can 
estimate the loss of condensate atoms by the scattering rate 
\be\label{eq:Ndothigh}
\dot N_0 = -\sum_{\nu>0}\Gamma_\nu .
\ee
This form of Floquet evaporative cooling leads to a dramatic reduction of heating. 

Within the local density approximation (\ref{eq:GammaLDA}), the loss rate of condensate atoms, given either by (\ref{eq:Ndotlow}) or by (\ref{eq:Ndothigh}), scales like $N_0^{7/5}$ with the number of condensate atoms $N_0$. This allows us to introduce the particle-number independent loss rate $\kappa$ defined by $\dot{N}_0=-\kappa N_0^{7/5}$.

\subsection{Resonant scattering in the Floquet picture} 

\subsubsection{Heating processes in the Floquet picture}
A periodically driven system with Hamiltonian 
$\Ho(t)=\Ho(t+T)=\sum_m \Ho^{(m)} e^{im\omega t}$
possesses quasistationary solutions of the time-dependent Schr\"odinger equation
called Floquet states,
\be
|\psi_n(t)\ra = e^{-i\varepsilon_n t/\hbar}|u_n(t)\ra 
= e^{-i\varepsilon_{nm} t/\hbar}|u_{nm}(t)\ra.
\ee
where $\varepsilon_{nm}=\varepsilon_n+m\hbar\omega$ and
$|u_{nm}(t)\ra=|u_{nm}(t+T)=e^{im\omega t}|u_{nm}(t)\ra$ denote the quasienergies
and the Floquet modes. The integers $m$ describe the fact that quasienergies are 
defined modulo $\hbar\omega$ only. From the time-dependent Schr\"odinger equation,
we obtain the equation
\be\label{eq:modulo}
\big[\Ho(t)-i\hbar\partial_t\big]|u_{nm}(t)\ra 
=\varepsilon_{nm}|u_{nm}(t)\ra,
\ee
which defines an eigenvalue problem in the extended Floquet Hilbert space 
\cite{Sambe73}. A complete basis of this space is given by the states
$|\alpha, m\rra$, representing a time-periodic state $|\alpha\ra e^{im\omega t}$ 
in the (standard) state space of the system. Here the states $|\alpha\ra$ form a 
complete basis of the lattice system and the Fourier index $m$ can assume any 
integer. The scalar product between two states $|u\rra$ and $|v\rra$, which 
represent time periodic states $|u(t)\ra=|u(t+T)\ra$ and $|v(t)\ra=|v(t+T)\ra$, 
is defined as
\be
\lla u|v\rra = \frac{1}{T}\int_0^T \! \rd t\, \la u(t)|v(t)\ra.
\ee
Marking operators by an overbar when acting in Floquet space, the quasienergy 
operator $\Qo(t)=\Ho(t)-i\hbar\partial_t$, which plays the role of a
time-independent Hamiltonian, possesses the matrix elements 
\be
\lla\alpha',m'|\bar{Q}|\alpha, m\rra 
= \frac{1}{T}\int_0^T \! \rd t\, e^{-i(m'-m)\omega t}
\la\alpha'|\big[\Ho(t)-i\hbar\partial_t\big]|\alpha\ra
\ee
so that
\be\label{eq:Q}
\lla\alpha',m'|\bar{Q}|\alpha, m\rra = \la\alpha'|\Ho^{(m'-m)}|\alpha\ra+\delta_{m'm}m\hbar\omega .
\ee
This structure guarantees the $\hbar\omega$-periodic quasienergy spectrum
(\ref{eq:modulo}). It resembles the problem of a quantum system described by the 
Hamiltonian $\Ho^{(0)}$ coupled to a photon-like mode, with $m$ playing the role
of the photon number relative to a large background occupation. 

By finding a unitary operator $\bar{U}_F$ that block diagonalizes the
quasienergy operator with respect to $m$,
\be
\lla\alpha',m'|\bar{U}^\dag_F\bar{Q}\bar{U}_F|\alpha, m\rra 
= \delta_{m'm} \Big[\la\alpha'|\Ho_F|\alpha\ra+m\hbar\omega \Big],
\ee
the problem of the driven system can be reduced to that of the time-independent 
effective Hamiltonian $\Ho_F$. This procedure is equivalent to performing a gauge
transformation with a time-periodic unitary operator $\Uo_F(t)=\Uo(t+T)$ in the 
original state space that leads to a time-independent Hamltonian. 

The basic idea of Floquet engineering is to design a driving protocol $\Ho(t)$
so that the effective Hamiltonian $\Ho_F$ acquires desired properties. For a 
driven many-body system, however, it is typically not possible to compute the
effective Hamiltonian exactly. Moreover, from the point of view of quantum 
simulation, we rather wish to effectively realize simple models with well-defined 
properties, whereas the exact Floquet Hamiltonian will typically be very complex 
object. For these reasons, one usually computes the effective Hamiltonian 
approximately. A common strategy is to employ a gauge transformation
described by a time-periodic unitary operator $\Uo(t)=\Uo(t+T)$, so that in the 
rotating frame the Hamiltonian $\Ho'(t)=\sum_m \Ho^{\prime(m)} e^{im\omega t}$
can be approximated by its time average $\Ho'(t)\approx\Ho^{\prime(0)}$. In
Floquet space, such a rotating-wave approximation corresponds to neglecting 
the off-diagonal blocks $(m'\ne m)$ of the quasienergy operator. 

Taking into account these off-diagonal terms gives rise to two different types of 
corrections: (i) The first type of correction comprises transitions and dephasing 
between unperturbed states of the same subspace $m$, result from the perturbative 
admixture of states from different subspaces. This effect can be captured by 
computing corrections to the unperturbed effective Hamiltonian (e.g. using a
high-frequency expansion \cite{GoldmanDalibard14, EckardtAnisimovas15}) and shall 
not be considered as heating. (ii) The second type of corrections are transitions 
between two (almost) degenerate unperturbed states of different subspaces $m$ and
$m'\ne m$. These processes correspond to resonant transitions, where the energy
of the Floquet engineered system changes by integer multiples $(m'-m)\hbar\omega$ 
of the photon energy $\hbar\omega$. They cannot be captured by adding time-independent corrections to the unperturbed time-independent 
effective Hamiltonian $\Ho^{\prime(0)}$ and must be considered as heating (if they
increase the energy). These are the processes to be investigated here. For that
purpose, we will employ perturbation theory in Floquet space.

\subsubsection{Estimating heating from perturbation theory}

Let $|n\ra$ be the eigenstates of $\Ho^{\prime(0)}$ with energy $\varepsilon_n$, 
\be
\Ho^{\prime(0)}|n\ra=\varepsilon_n|n\ra,
\ee
and $|n,m\rra$ the corresponding basis states in Floquet space. Within the
rotating-wave approximation 
$|n,m\rra$ is an eigenstate of the quasienergy operator with quasienergy 
$ \varepsilon_{nm}=\varepsilon_n+m\hbar\omega$.
Let us, moreover, assume that initially the system is prepared in the ground
state $|0\ra$ of $\Ho^{\prime(0)}$. Then heating requires that there are excited 
states $n$ that, modulo $\hbar\omega$, are near degenerate to the ground state, 
\be
\varepsilon_n = \varepsilon_0+\nu\hbar\omega +\delta
\ee
with $\nu>0$. The detuning $|\delta|$ has to be small compared to the matrix 
matrix element $C^{(\nu)}_{n}$ that couples the state $|0,m\rra$ to the state 
$|n, m-\nu\rra$. 

Such a coupling matrix element can either be given directly by a matrix element 
of the quasienergy operator
\be\label{eq:C1st}
C^{(\nu) 1\text{st}}_{n}=\lla n,m|\bar{Q}|0,0\rra = \la n|\Ho^{\prime(m)}|0\ra
\quad\text{with}\quad
m=-\nu.
\ee 
It can, however, also result from indirect coupling processes via virtual 
intermediate states. In this case it can be obtained using degenerate 
perturbation theory. In $p$th order it describes processes via $p-1$ 
intermediate states $|n_i, m_i\rra$. A second-order coupling matrix element takes 
the form 
\be\label{eq:C2nd}
C^{(\nu) 2\text{nd}}_{n}={\sum_{n_1m_1}}^{\!\prime}
\frac{\la n|\Ho^{\prime(m-m_1)}|n_1\ra\la n_1|\Ho^{\prime(m_1)}|0\ra}
{\varepsilon_0-(\varepsilon_{n_1}+m_1\hbar\omega)},
\ee
where we have assumed that $\delta$ can be neglected relative to the energy in 
the denominator. The prime at the sum shall indicate that we are only summing over 
states that are energetically well separated from the group of almost degenerate 
states to which $|0,0\rra$ couples resonantly. 

Generally, a $p$th-order process is 
described by a matrix element of the order of
\bes\label{eq:Cmth}
&&C^{(\nu) p\text{th}}_{n}\sim
\\\nonumber&&
{\sum_{\{n_im_i\}}}^{\!\!\!\prime}\frac{\la n|\Ho^{\prime(m-m_{p-1})}|n_{p-1}\ra
	\cdots\la n_1 |\Ho^{\prime(m_1)}|0\ra}
{(\varepsilon_0-\varepsilon_{n_{p-1}}-m_{p-1}\hbar\omega)\cdots
	(\varepsilon_0-\varepsilon_{n_1}-m_1\hbar\omega)}.
\ees

\subsubsection{General form of the Hamiltonian}
In the following, we will introduce two different rotating frames, depending on 
whether we are considering the lower or the larger driving frequency. For each of them the Hamiltonian $\Ho'(t)$ can be decomposed like
\be\label{eq:Hdec}
\Ho'(t) = \Ho_\text{bs} + \sum_{m} \big[\Ho^{(m)}_\text{sp} + \Ho^{(m)}_\text{tp}\big]e^{im\omega t}.
\ee
Here $\Ho_\text{bs}$ is time-independent and describes the single-particle band
structure
\be
\Ho_\text{bs} = \sum_{b \bk} \varepsilon'_b(\bk) \aa_{b\bk}\ao_{b\bk},
\ee
where $\ao_{b\bk}$ is the bosonic annihilation operator for a boson in the 
single-particle state $|b\bk\ra'$ characterized by the band index $b$ and
the (quasi)momentum $\bk$. The corresponding energy reads
\be
\varepsilon'_b(\bk) = \varepsilon'_b(k_x) + E_\perp(\bk_\perp).
\ee
All other single-particle terms are collected in the terms
\be\label{eq:AmH}
\Ho_\text{sp}^{(m)} = \sum_{\bk}\sum_{b' b} A_{b'b,\bk}^{(m)}
\aa_{b'\bk}\ao_{b\bk} .
\ee
The interactions are contained in the terms
\be\label{eq:Bm}
\Ho_\text{tp}^{(m)} = \sum_{\{b\bk\}} B_{\{b\bk\}}^{(m)}
\aa_{b_4\bk_4}\aa_{b_3\bk_3}\ao_{b_2\bk_2}\ao_{b_1\bk_1},
\ee
having matrix elements $B_{\{b\bk\}}^{(m)}$ that vanish unless (quasi)momentum
is conserved. 

For our analysis of heating channels, we will neglect the impact of the ordinary 
zero-photon interactions $\Ho^{(0)}_\text{tp}$ on the eigenstates of
$\Ho^{\prime(0)}$. 
(However, we do take these interactions into account, in order to compute the 
condensate's density distribution in the trap, as described in
section\ref{sec:trap} above). The effect of interactions is to induce quantum 
fluctuations that cause a slight depletion of the condensate and make the 
dispersion relation phonon-like for small $\bk$, as it is described by Bogoliubov 
theory. However, for weak interactions both have no major impact on the heating
rates. Within this approximation the eigenstates of $\Ho^{\prime(0)}$ are given
by Fock states $|\bn \ra$ characterized by the vector $\bn$ of occupation numbers
$n_{b\bk}$. The energy of this state is given by
$\varepsilon_{\bn}=\sum_{b\bk} n_{b\bk} \varepsilon'_b(\bk)$.
The corresponding Floquet-Fock states are denoted $|\bn, m\rra$; their unperturbed 
quasienergies read $\varepsilon_{\bn m}(\bk) = \varepsilon_{\bn} + m\hbar\omega$.

The ground state $|0\ra$ is given by the condensate state with all particles 
occupying the condensate mode $|b=0,\bk=\bq\ra'$. For our analysis, we 
moreover, consider Fock states with a single particle occupying the excited state
$|b,\bk\ra'$ and with two particles occupying the excited states $|b_1,\bk_1\ra'$ 
and $|b_2,\bk_2\ra'$, while all other particles remain in the condensate. These 
states shall be denoted by $|1(b,\bk)\ra$ and $|2(b_1,\bk_1;b_2,\bk_2)\ra$, 
respectively. 
The matrix element for resonant $\nu$-photon scattering
$C^{(\nu)}_{b'b}(\bk,\bq)$ describes the coupling of the Floquet-Fock state
$|0,0\rra$, denoting the $m=0$ ground state, with the Floquet-Fock state
$|2(b,\bq+\bk;b',\bz-\bk), -\nu\rra$, denoting a state with two additional excited
particles, but $\nu$ photons less.

\subsection{Heating rates for driving frequencies below the band gap}

\subsubsection{The rotating frame}
In the regime where the driving frequency is well below the gap separating the 
lowest from the first excited band discussed in section \ref{sec:sp_low}, it is 
convenient to describe the system in a reference frame that is translated with
respect to quasimomentum by $\bq(t)=q(t){\bm e}_x$, as defined in
Eq.~(\ref{eq:q(t)}), while the band index is conserved. This is accomplished by 
the transformation (\ref{eq:U1}) and leads to the single-particle Hamiltonian
reads (\ref{eq:h'}).
As long as interactions are happen on site, which is a good approximation for the 
lowest band(s), the interaction Hamiltonian (\ref{eq:Hint_bk}) is not altered by 
the shift $\bq(t)$ in quasimomentum.
The time-dependence of the energies $E_b\big(\bk+\bq(t)\big)$ appearing
in Eq.~(\ref{eq:h'}) is contained in the band energies $E_b(k_x+q(t))$ that we 
Fourier-decompose like
\be
E_b(k_x+q(t)) =\sum_{m=-\infty}^\infty E^{(m)}_b(k_x) e^{im\omega t}.
\ee
Assuming that for the lowest band(s) the kinetics of the system is governed by 
nearest-neighbor tunneling only so that the dispersion relation is given by
Eq.~(\ref{eq:Eb}), we obtain
\bes
E^{(m)}_b(k_x) &=& \frac{1}{T}\int_0^T \!\rd t\, e^{-im\omega t}
\big[ \varepsilon_b -2J_b\cos(ak_x- \alpha\sin\big(\omega t)\big)\big]
\nonumber\\&=&   
\varepsilon_b \delta_{m,0} -J_b \mathcal{J}_m(\alpha)\Big[(-)^m e^{iak_x}+                  e^{-iak_x}\Big].
\ees

In order to get rid of the time-dependence of the diagonal terms in
Eq.~(\ref{eq:h'}), let us perform a second gauge transformation, where we 
integrate out the time-periodic part $\sum_{m\ne0} E^{(m)}_b(k_x) e^{im\omega t}$ 
of the band energies $E_b\big(\bk+\bq(t)\big)$. For that purpose we employ the 
unitary operator
\be\label{eq:U1'}
\Uo_1'(t) = \exp\bigg[i \sum_{b\bk}\sum_{m\ne0}\chi^{(m)}_{b\bk}e^{im\omega t}
\aa_{b\bk}\ao_{b\bk}\bigg]
\ee
with
\be\label{eq:chi1}
i\chi_{b\bk}^{(m)}
    = -\frac{E^{(m)}_b(k_x)}{m\hbar\omega}
    = \frac{\mathcal{J}_m(\alpha)}{m}D_{b\bk}^{(m)},
\ee
and
\be\label{eq:D}
D_{b\bk}^{(m)} 	= \frac{2J_b}{\hbar\omega}\left\{ \begin{array}{ll} 
	\cos(ak_x), & \text{ even }|m|\\
	\frac{1}{i}\sin(ak_x), & \text{ odd }|m| \,.\\
\end{array}
\right. 
\ee

The transformed Hamiltonian can be evaluated using
$\exp(-i\chi\aa\ao)\hat{a}\exp(i\chi\aa\ao)=\exp(i\chi)\hat{a}$. It is of the
form (\ref{eq:Hdec}). The first term describes a renormalized band structure,
\be
\Ho_\text{bs} = \sum_{\bk b}\varepsilon_b(\bk)\aa_{b\bk}\ao_{b\bk}
\ee
with dispersion relation
\be
\varepsilon_b(\bk) =  \varepsilon_b -2J_b\mathcal{J}_0(\alpha) \cos(ak_x)
+ E_\perp(\bk_\perp) .
\ee
The second term describes single-particle interband transitions,
\be\label{eq:Hsp1}
\Ho_\text{sp}(t)= K \sum_{\bk}\sum_{b'b}\eta_{b'b} \cos(\omega t)
e^{-i[\chi_{b'\bk}(t)-\chi_{b\bk}(t)]}
\aa_{b'\bk}\ao_{b\bk} .
\ee
Finally, the third term comprises two-particle scattering processes,
\bes\label{eq:Htp1}
\Ho_\text{tp}(t) &=& \frac{g}{2 L^3} {\sum_{\{b\bk\}}}' \zeta_{\{b\}}
e^{-i[\chi_{b_4\bk_4}(t)+\chi_{b_3\bk_3}(t)
	-\chi_{b_2\bk_2}(t)-\chi_{b_1\bk_1}(t)]}
\nonumber\\ && \qquad\quad\times
\aa_{b_4\bk_4}\aa_{b_3\bk_3}\ao_{b_2\bk_2}\ao_{b_1\bk_1} .
\ees

Let us now Fourier-decompose these terms and, thus, determine the coefficients 
appearing in Eqs.~(\ref{eq:AmH}) and (\ref{eq:Bm}). In order to obtain clear and 
simple expressions, we will write down only the most dominant contribution to 
each Fourier component only. When expanding the exponential functions appearing in 
Eqs.~(\ref{eq:Hsp1}) and (\ref{eq:Htp1}) with respect to their argument 
and comparing the result to their Fourier decomposition,
\be
\exp\Big({i\chi_{b\bk}(t)}\Big) 
= \sum_{\nu=0}^\infty\frac{1}{\nu!}\Big({i\sum_{m\ne 0} \chi_{b\bk}^{(m)}e^{im\omega t}}\Big)^\nu 
= \sum_{\mu=-\infty}^\infty\!\!e_{b\bk}^{(\mu)}e^{i\mu\omega t},
\ee
we can identify the leading contributions to each Fourier component
$e_{b\bk}^{(\mu)}$ to be given by
\be
e_{b\bk}^{(0)} \simeq 1, \qquad e_{b\bk}^{(\mu)} \simeq i\chi_{b\bk}^{(\mu)}
\qquad\text{for}
\qquad \mu\ne0.
\ee
Namely, products of several $\chi_{b\bk}^{(\mu)}$, as they appear for $\nu\ge2$, 
are not relevant, since each $\chi_{b\bk}^{(\mu)}$ contributes a factor
$J_b/(\hbar\omega)\ll1$. 
Note, moreover, that with respect to its argument the Bessel function
appearing in Eq.~(\ref{eq:chi1}) for $i\chi_{b\bk}^{(\mu)}$ behaves like 
\be
\mathcal{J}_m(\alpha)\simeq \frac{[\sgn(m)]^{|m|}}{|m|!}\Big(\frac{\alpha}{2}\Big)^{|m|}
\simeq \frac{1}{\sqrt{2\pi|m|}}\bigg(\frac{e\alpha}{2|m|}\bigg)^{|m|}
\ee
for small $\alpha$ (second expression) and large $|m|$ (third expression).  
Thus, $\chi_{b\bk}^{(m)}$ is exponentially suppressed for large $|m|$ with
$|m|>(e/2)\alpha\approx1.36\alpha$.

We can now write down the leading contribution to the Fourier components
$\Ho_\text{sp}^{(m)}$ and $\Ho_\text{tp}^{(m)}$. Taking into account also the 
factor $\cos(\omega t)=\frac{1}{2}\big[e^{i\omega t}+e^{-i\omega t}\big]$ 
appearing in $\Ho_\text{sp}(t)$, for the former we find the coefficients
\be
A^{(0)}_{b'b,\bk} 
\simeq
\eta_{b'b}\mathcal{J}_1(\alpha)\alpha\hbar\omega
\big[D_{b\bk}^{(1)}- D_{b'\bk}^{(1)}\big],
\ee
and 
\be\label{eq:A1}
A^{(\pm 1)}_{b'b,\bk} \simeq \frac{1}{2}\alpha\hbar\omega\eta_{b'b},
\ee
as well as 
\be\label{eq:Am}
A^{(\pm|m|)}_{b'b,\bk}
\simeq
\eta_{b'b}\frac{(\pm)^{|m|}\mathcal{J}_{|m|-1}(\alpha)\alpha\hbar\omega}{2(|m|-1)}
\Big[D_{b'\bk}^{(|m|-1)}-D_{b\bk}^{(|m|-1)}\Big]
\ee
for $|m|\ge2$. 
Thus, while we have the scaling
$ A^{(\pm |m|)}_{b'b,\bk}\propto (J_{b'}-J_b)\eta_{b'b}\alpha^{|m|}/|m|!$ for
$|m|\ge2$, the $m=\pm1$ coefficients (\ref{eq:A1}) are a factor of
$\hbar\omega/(J_{b'}-J_b)\gg1$ larger.

For the two-particle processes the coefficients read
\be
B^{(0)}_{\{b\bk\}} \simeq \frac{g\zeta_{\{b\}}}{L^3},
\ee			
so that $\Ho_\text{tp}^{(0)} \simeq \Ho_\text{int}$,
as well as
\be\label{eq:B}
B^{(m)}_{\{b\bk\}} 
\simeq
\frac{g\zeta_{\{b\}}}{2L^3} \frac{\mathcal{J}_m(\alpha)}{m}
\big[D_{b_1\bk_1}^{(m)}+D_{b_2\bk_2}^{(m)}-D_{b_3\bk_3}^{(m)}-D_{b_4\bk_4}^{(m)}
\big]
\ee
for $|m|\ge1$. Therefore, zero-photon scattering is much faster than 
resonant $m$-photon scattering with $m\ne0$.

\subsubsection{Heating rate}
According to the resonance condition (\ref{eq:condition}), for driving frequencies 
well below the band gap, scattering processes into the first (or even a higher)
excited band, correspond to $\nu$ photon processes with large $\nu$, which are 
strongly suppressed with respect to resonant scattering processes within the 
lowest band. Thus, the dominant channel for interaction-induced heating is
\emph{resonant intraband scattering}, where a pair of condensate atoms scatters 
from the condensate mode $|0\bz\ra$ into the excited Bloch states 
$|0\bk\ra$ and $|0-\bk\ra$, so that (quasi)momentum is conserved. The dominant 
processes would be single-photon scattering. However, 
the corresponding coupling matrix element, which is of the form (\ref{eq:C1st}),
vanishes by symmetry for odd photon numbers $\nu$ as long as the condensate  
possesses quasimomentum $q=0$. 
Thus, the leading process will be given by two-photon resonant scattering
generated by $\Ho^{(-2)}_\text{int}$.
The matrix element for resonant scattering with even $\nu$ is of the form 
(\ref{eq:C1st}) and reads
\bes
C^{(\nu)}_{00}(\bk) &=& \la 2(0\bk,0-\bk)|\Ho^{(-\nu)}_\text{tp}|0\ra
= NB^{(-\nu)}_{0\bk,0-\bk,0\bz,0\bz} 
\nonumber\\
&\simeq&- \frac{Ng}{L^3}\frac{\zeta J_0}{\hbar\omega}
\frac{\mathcal{J}_\nu(\alpha)}{\nu/2}[1-\cos(ak_x)]
\quad\text{for even $\nu$} .
\nonumber\\
\ees

In order to compute the heating rate, let us first evaluate the dimensionless 
scattering rate (\ref{eq:gamma}). As long as $\nu\hbar\omega$ is larger than twice 
the width of the dressed lowest band, $8J_0\mathcal{J}_0(\alpha)$, all modes $k_x$
contribute to the integral and we find
\be
\gamma_\nu 
= \bigg(\frac{2\zeta J_0\mathcal{J}_\nu(\alpha)}{\nu\hbar\omega}\bigg)^2
\underbrace{
	\frac{a}{2\pi}\int_{-\pi/a}^{\pi/a}\!\rd k_x\, [1-\cos(ak_x)]^2}_{3/2}
\ee
For lower driving frequencies, when 
$s_\nu \equiv \nu\hbar\omega/[8J_0\mathcal{J}_0(\alpha)] < 1 $,
the resonance condition (\ref{eq:condition}) is fulfilled only for modes
$|k_x|<k_\text{max}$ with $[1-\cos(a k_\text{max})]=2s_\nu$. Integrating over 
the contributing modes, we define
\bes
g(s_\nu)
&=&\frac{2}{3}\frac{a}{2\pi}\int_{-k_\text{max}}^{k_\text{max}}\!\rd k_x\, [1-\cos(ak_x)]^2
\nonumber\\
&=&    \bigg[\frac{1}{2}-\frac{1}{\pi}\arcsin(1-2s_\nu)\bigg] 
-\frac{2s_\nu+6}{3\pi}\sqrt{s_\nu(1-s_\nu)}     .
\nonumber\\
\ees
With that and the convention
$ g(s_\nu)=1\,\text{for}\, s_\nu\ge1$,
the dimensionless scattering rate is given by
\be\label{eq:gamma_low}
\gamma_\nu 
= 6 v_\nu g(s_\nu)\bigg(\frac{\zeta J_0\mathcal{J}_\nu(\alpha)}{\nu\hbar\omega}\bigg)^2,
\ee
where $v_\nu=1$ ($v_\nu=0$) for even (odd) $\nu$.

The experimental measurements show that the condensate possesses a finite 
width in momentum space. As a result of this finite width also scattering 
processes corresponding to an odd photon number, and in particular single-photon 
processes, acquire finite matrix elements. Assuming the condensate to be formed 
in a coherent superposition over several quasimomenta $k_x$ of width $\Delta k_x$
around $k_x=0$, we find scattering rates are still of the form
(\ref{eq:gamma_low}), but now $v_\nu$ takes two different values
$v_o$ and $v_e$ for odd and even $\nu$, respectively. We have computed these 
factors for the experimentally measured width $\Delta k_x$ and took them into account in the data presented in the main text.

\subsection{Heating rates for driving frequencies above the band gap}

\subsubsection{The rotating frame}

Let us now consider the regime where the driving frequency is larger than the 
gap separating the lowest from the first excited band. As has been pointed out
in section \ref{sec:sp_high}, it is convenient to describe the system in 
a reference frame that is translated with respect to momentum by
$\bq(t)=q(t){\bm e}_x$. This is accomplished by the 
transformation (\ref{eq:U2})
and gives rise to the single-particle Hamiltonian (\ref{eq:h''}), which we
write in second quantization as
\be
\Ho_0''(t) = \sum_{\bk b} E_b(\bk)\aa_{b\bk}\ao_{b\bk}
+\sum_{\bk b'b} V_{b'b}(k_x,t) \aa_{b'\bk}\ao_{b\bk}.
\label{eq:H0''}
\ee
Here $\aa_{b\bk}$ creates a particle in the 
undriven single-particle Bloch state $|b\bk\ra$ with undriven energy
$E_b(\bk)$. The driving is captured by the matrix elements 
\be
V_{b'b}(k_x,t)=\frac{\hbar^2}{2m}\big[2q(t)p_{b'b}(k_x)
+q^2(t)\delta_{b'b}\big],
\ee
where
\be
p_{b'b}(k_x) = \la b' k_x|-i\partial_x|bk_x \ra 
=  k_x\delta_{b'b} +\frac{2\pi}{a} \beta_{b'b}(k_x)
\ee
denote the matrix elements of the momentum (wave number) with respect to the
Bloch states $|b\bk\ra$. Here we introduced
$\beta_{b'b}(k_x) =\sum_\beta u_{b'\beta}^*(k_x)\beta u_{b\beta}(k_x)$,
with  $u_{b\beta}(k_x)=\la\beta k_x|b k_x\ra$.
The interaction Hamiltonian (\ref{eq:Hint_bk}) is not altered by the gauge 
transformation.

It is convenient to decompose the Hamiltonian (\ref{eq:H0''}) into diagonal
and band-coupling terms.
The time-dependent diagonal energies 
\bes
E_b(\bk,t) &=& E_b(\bk) +\frac{\hbar^2}{2m} \big[2 q(t) p_{bb}(k_x)+ q^2(t)\big]
\nonumber\\
&=& \sum_{m=-2}^2 E_b^{(m)}(\bk) e^{im\omega t}, 
\ees
possess Fourier components
\bes
E_b^{(0)}(\bk) &=& E_b(k_x) + E_\perp(\bk_\perp) + \frac{\alpha^2}{2\pi^2}E_r
\\
E_b^{(\pm1)}(\bk) &=&\pm  \frac{2\alpha}{i\pi} E_r 
\Big[\frac{ak_x}{2\pi}+\beta_{bb}(k_x)\Big]
\\
E_b^{(\pm2)}(\bk) &=& -  \frac{\alpha^2}{4\pi^2}E_r.
\ees

In order to get rid of the time-dependence of the diagonal terms
$\propto \aa_{b\bk}\ao_{b\bk}$ in Eq.~(\ref{eq:H0''}), let us
again perform a second gauge transformation, where we integrate out the
time-periodic part $\sum_{m\ne0} \varepsilon^{(m)}_b(k_x) e^{im\omega t}$ of the 
band energies $E_b\big(\bk+\bq(t)\big)$. For that purpose we again employ a 
unitary operator of the form (\ref{eq:U1'}) and (\ref{eq:chi1}).
The transformed Hamiltonian can be decomposed into three terms. 
The first term of the transformed Hamiltonian 
describes the unperturbed band structure,
\be
\Ho_\text{bs} = \sum_{\bk b}\varepsilon_b(\bk)\aa_{b\bk}\ao_{b\bk}.
\ee
The second term describes single-particle interband transitions,
\be\label{eq:Hsp}
\Ho_\text{sp}(t)= -\,\frac{4\alpha}{\pi} E_r\sin(\omega t)
\sum_{\bk}\sum_{b'b}\beta_{b'b}(k_x) e^{i\theta_{b'b}(k_x,t)}
\aa_{b'\bk}\ao_{b\bk} ,
\ee
with time-dependent phases 
\bes
\theta_{b'b}(k_x,t) &=& \chi_{b\bk}(t)-\chi_{b'\bk}(t)
\nonumber\\
&=& \frac{4\alpha}{\pi}\frac{E_r}{\hbar\omega}
\Big[\beta_{bb}(k_x)-\beta_{b'b'}(k_x)\Big]\cos(\omega t)
\nonumber\\ &\equiv& 2\theta^{(1)}_{b'b}(k_x) \cos(\omega t).
\ees
Finally, the interaction term transforms into
\be\label{eq:Htp}
\Ho_\text{tp}(t) = \frac{g}{2L^3} {\sum_{\{b\bk\}}}' \zeta_{\{b\bk\}}
e^{i\theta_{\{b\bk\}}(t)}
\aa_{b_4\bk_4}\aa_{b_3\bk_3}\ao_{b_2\bk_2}\ao_{b_1\bk_1} ,
\ee
with time-dependent phases
\bes
\theta_{\{b\bk\}}(t)&=& \chi_{b_1\bk_1}(t)+\chi_{b_2\bk_2}(t)
-\chi_{b_3\bk_3}(t)  -\chi_{b_4\bk_4}(t)
\nonumber\\&=&
\frac{4\alpha}{\pi}\frac{E_r}{\hbar\omega}\cos(\omega t)
\Big[k_{x1}+k_{x2}-k_{x3}-k_{x4}
\nonumber\\&&               
+ \beta_{b_1b_1}(k_x)+\beta_{b_2b_2}(k_x)
-\beta_{b_3b_3}(k_x)-\beta_{b_4b_4}(k_x)\Big]
\nonumber\\ &\equiv& 2\theta^{(1)}_{\{b\bk\}} \cos(\omega t).
\ees

By Fourier-decomposing the $\Ho_\text{sp}(t)$ and $\Ho_\text{tp}(t)$
we obtain the coefficients defined in Eqs.~(\ref{eq:Am}) and (\ref{eq:B}),
\bes
A^{(m)}_{b'b}(\bk) &=& -(i)^m\frac{2\alpha}{\pi}\beta_{b'b}(k_x)E_r
\nonumber\\&&\times\,
\Big[\mathcal{J}_{m-1}\big(2\theta^{(1)}_{b'b}(k_x)\big)
+ \mathcal{J}_{m+1}\big(2\theta^{(1)}_{b'b}(k_x)\big)\Big]
\nonumber\\
\ees
and 
\be
B^{(m)}_{\{b\bk\}} = (i)^m\frac{g}{2L^3}\zeta_{\{b\bk\}}
\mathcal{J}_{m}\big(2\theta^{(1)}_{\{bk_x\}}\big).
\ee

The phases $\theta^{(1)}$ appearing in the arguments of the Bessel functions are
small, since they scale like $\alpha E_r/\hbar\omega$ with respect to driving 
frequency and strength. We will therefore, in a first approximation, consider
only the leading terms. These involve the Bessel functions of order zero, which 
behave like
$\mathcal{J}_0(2\theta^{(1)})\simeq (2\theta^{(1)})^0=1$ for small
$\theta^{(1)}$. In this order, the only relevant matrix elements are given by
\be\label{eq:A_1}
A^{(\pm1)}_{b'b}(\bk) \simeq -E_r\frac{2\alpha}{\pi}\beta_{b'b}(k_x),
\ee
describing single-photon single-particle interband transitions as well as by
\be\label{eq:B0}
B^{(0)}_{\{b\bk\}} \simeq \frac{g}{2L^3}\zeta_{\{b\bk\}},
\ee 
describing zero-photon intra- and inter-band scattering processes.

The leading corrections involve terms that scale linearly with respect to
$\theta^{(1)}$. The corresponding coefficients read
\be\label{eq:A2}
A^{(\pm2)}_{b'b}(\bk) 
\simeq \pm \frac{E_r^2}{\hbar\omega}\left(\frac{2\alpha}{\pi}\right)^2
\beta_{b'b}(k_x)[\beta_{b'b'}(k_x)-\beta_{bb}(k_x)],
\ee
while $A^{(0)}_{b'b}(\bk) = 0$, and 
\bes\label{eq:B1}
B^{(\pm1)}_{\{b\bk\}} &&\simeq \pm i \frac{g}{2L^3}\frac{E_r}{\hbar\omega}
\frac{2\alpha}{\pi}
\zeta_{\{b\bk\}} 
\big[k_{x1}+k_{x2}-k_{x3}-k_{x4}
\nonumber\\&&                                  
+\, \beta_{b_1b_1}(k_{x1})+\beta_{b_2b_2}(k_{x2})
- \beta_{b_3b_3}(k_{x3})-\beta_{b_4b_4}(k_{x4})\big].
\nonumber\\
\ees

\subsubsection{Heating rate}
We will compute the matrix elements for an $\nu$ photon process using degenerate
perturbation theory. Assuming a driving frequency of
$\hbar\omega/E_r=7.5$ which lies in the second band gap (so that neither the first 
nor the second band are directly coupled to the lowest band
via momentum-conserving single-photon processes), we will take into account the
most dominant processes only. These are determined according to the following 
three principles:
\begin{itemize} 
	\item[(i)] We will consider only terms involving the dominant matrix elements
	(\ref{eq:A_1}) and (\ref{eq:B0}), describing single-photon interband 
	transitions and zero-photon scattering, respectively.  
	\item[(ii)] Assuming that scattering processes are slow compared to single-
	particle interband transitions, $g N/L^3\ll E_r\alpha$, in $n$th order 
	perturbation theory, we will take into account a single scattering process 
	and $n-1$ single-particle inter-band processes. 
	\item[(iii)] We will take into account only perturbative contributions 
	involving virtual intermediate states that are separated in quasienergy by a 
	distance much smaller than $\hbar\omega$. 
\end{itemize}

\begin{figure}[t]
	\includegraphics[width=0.7\linewidth]{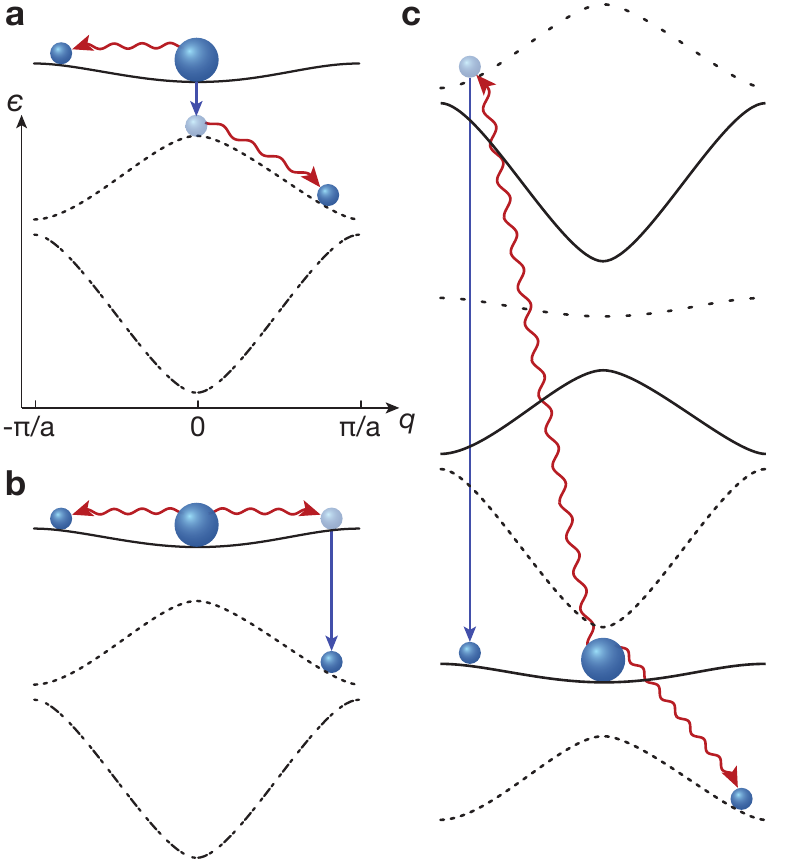}
	\centering
	\caption{\label{fig:C01} (a,b) Dominant single-photon second-order processes
		contributing to the matrix element $C^{(1)}_{10}(\bk_x)$. From top to 
		bottom, the three depicted bands correspond to the unperturbed Floquet 
		Bloch bands with energies $\varepsilon_{00}(k_x)=\varepsilon_0(k_x)$,
		$\varepsilon_{1-1}(k_x)=\varepsilon_1(k_x)-\hbar\omega$, and 
		$\varepsilon_{2-2}(k_x)=\varepsilon_2(k_x)-2\hbar\omega$. The
		Bose-Einstein condensate (BEC) is occupying the state with $b=0$, $\bk=0$,
		and $m=0$ (the center of the uppermost band that shown). The solid blue
		arrow symbolizes the single-particle single-photon process with matrix 
		element	$A^{(-1)}_{b'b,\bk}$. The pair of curved red arrows symbolizes a 
		two-particle zero-photon scattering processes with matrix element 
		$B^{(0)}_{b_4\bk,b_3-\bk,b_2\bz,b_1\bz}$. 
		(c) Subdominant process [violating the selection principle (iii)]. While 
		it gives rise to a matrix element that is small compared to those 
		associated with the	individual processes depicted in (a) and (b), it 
		still matters since the contributions from (a) and (b) interfere 
		destructively.}
\end{figure}

Following these rules, the leading heating processes appearing in second-order
perturbation theory are depicted in Fig.~\ref{fig:C01}(a) and (b). Here, from
top to bottom, the three bands depicted in both panels correspond to the
unperturbed Floquet-Bloch bands with energies
$\varepsilon_{00}(k_x)=\varepsilon_0(k_x)$, 
$\varepsilon_{1-1}(k_x)=\varepsilon_1(k_x)-\hbar\omega$, and 
$\varepsilon_{2-2}(k_x)=\varepsilon_2(k_x)-2\hbar\omega$. The blue arrow
symbolizes the single-particle single-photon process with matrix element
$A^{(-1)}_{b'b,\bk}$. The pair of curved red arrows stands for a two-particle
zero-photon scattering processes with matrix element
$B^{(0)}_{b_4\bq+\bk,b_3\bq-\bk,b_2\bq,b_1\bq}$. (In the figure, the condensate 
is assumed be formed in the $\bq=\bz$ mode). These terms contribute to the matrix 
element $C^{(1)}_{10}(\bk, \bq) $. The resonance condition (\ref{eq:condition})
determines how much energy has to be transferred into the transverse degree 
of freedom and with that the transverse momentum $\bk_\perp$,
\be
2E_\perp(\bk_\perp) = \hbar\omega-\varepsilon_1(q+k_x)-\varepsilon_0(q-k_x)
+2\varepsilon_0(q).
\ee
Note that the fact that $E_\perp(\bk_\perp)\ge0$ excludes processes involving
the matrix elements $A^{(1)}_{b'b,\bk}$ that raise the photon number.

\begin{figure}[t]
	\includegraphics[width=0.49\linewidth]{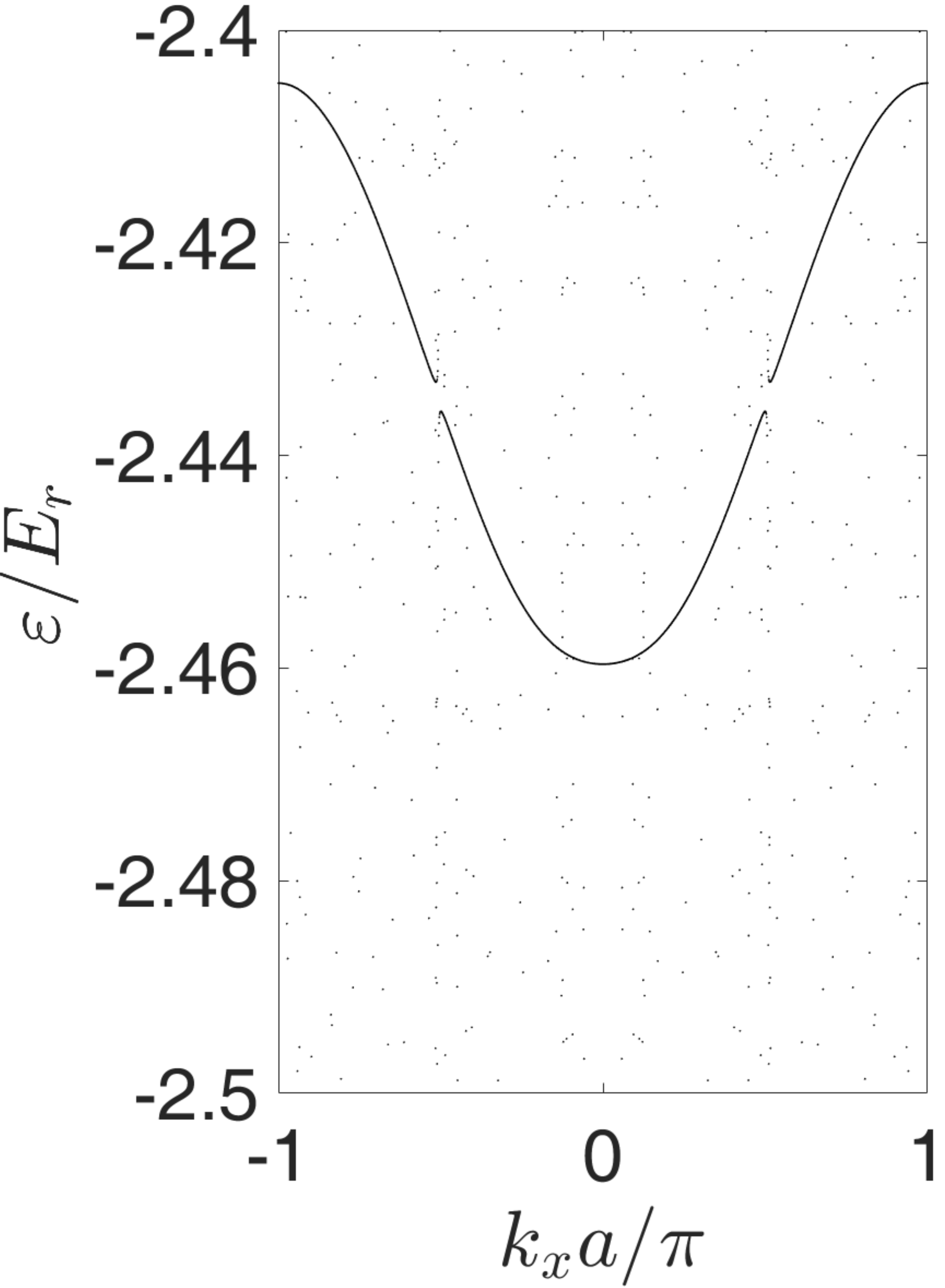}
	\includegraphics[width=0.49\linewidth]{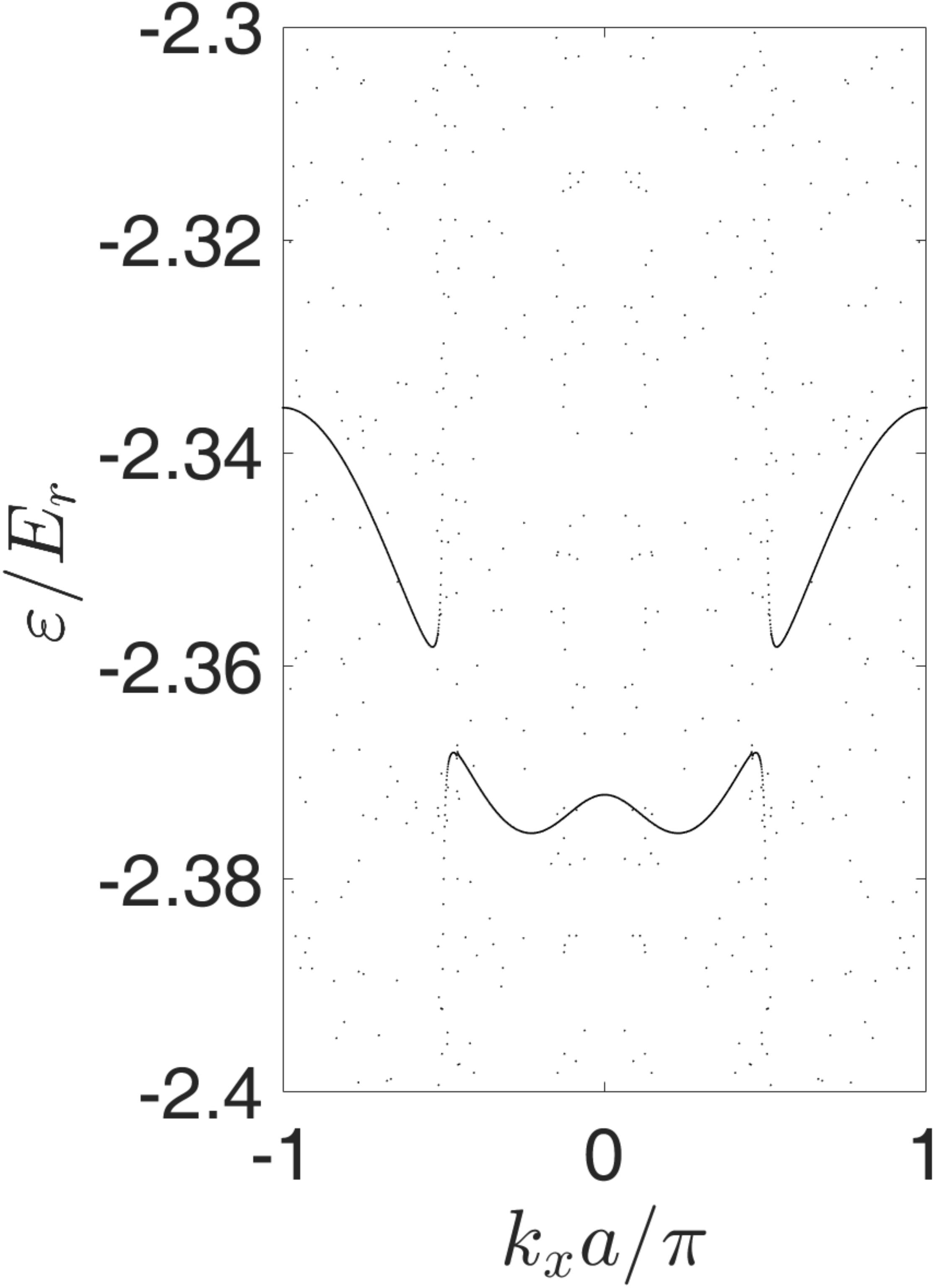}
	\centering
	\caption{\label{fig:DoubleWell} 
		Quasienergy band structure of the driven cosine lattice with $V_0/E_r=11$, 
		$\hbar\omega/E_r=7.5$, and $\alpha = 0.5$ (left panel)
		as well as $\alpha=1$ (right panel).} 
\end{figure}

The reason why we consider also the situation where the condensate is formed in
a state with finite quasimomentum $q$ is related to the fact that for $V_0/E_r=11$
and $\hbar\omega/E_r=7.5$ the quasienergy band emerging from the undriven ground 
band develops a double well structure for sufficiently strong driving amplitudes 
above $\alpha\approx0.75$ with two new minima at the quasimomenta $k_x=\pm q$. 
This effect is a result of the perturbative admixture of the much more dispersive 
$b=1$ band with one photon less (i.e. shifted by $-\hbar\omega$) to the ground 
band via single-particle coupling. This can be seen in the quasienergy spectra 
shown in Fig.~\ref{fig:DoubleWell}. In this situation time-of-flight measurements 
reveal that the condensate is displaced to one or both of the new minima of the 
ground band. With respect to the repulsive interactions it is favorable to occupy 
one of the minima only, so that the interpretation is that domains are formed 
where the condensate sits either in the state $q$ or $-q$. This behavior has also 
been measured in a recent experiment \cite{Parker_13}.

For the processes depicted in subfigure Fig.~\ref{fig:C01}(a) and (b), the virtual intermediate
quasienergies relative to the condensate energy $E_0(\bz)$ read
\bes
\varepsilon_{v}^{(a)} &=& \Delta_{10}(q)-\hbar\omega
\\
\varepsilon_{v}^{(b)} &=& \varepsilon_0(q+k_x)+\varepsilon_0(q-k_x)-2\varepsilon_0(q)+2E_\perp 
\nonumber\\
&=& \hbar\omega-\Delta_{10}(q+k_x),
\ees
with $\Delta_{b'b}(k_x) \equiv \varepsilon_{b'}(k_x)-\varepsilon_b(k_x)$.
The matrix elements of the two elementary processes involved in both terms 
multiply to the factors
\bes
M^{(a)} &=& \sqrt{N}^2A^{(-1)}_{10}(q)
\big[B^{(0)}_{1\bq+\bk,0\bq-\bk,1\bq,0\bq}
+B^{(0)}_{1\bq+\bk,0\bq-\bk,0\bq,1\bq} 
\nonumber \\ &&           
+\, B^{(0)}_{0\bq-\bk,1\bq+\bk,1\bq,0\bq}
+B^{(0)}_{0\bq-\bk,1\bq+\bk,0\bq,1\bq} \big]
\nonumber\\
\ees
and 
\bes
M^{(b)}&=&
\sqrt{N}^2A^{(-1)}_{10}(q+k_x) 
\nonumber\\&&
\times\,\big[B^{(0)}_{0\bq+\bk,0\bq-\bk,0\bq,0\bq}
+B^{(0)}_{0\bq-\bk,0\bq+\bk,0\bq,0\bq} \big],
\nonumber\\
\ees
so that the effective scattering matrix element, which is of the form 
(\ref{eq:C2nd}), reads
\bes\label{eq:C1ab}
C^{(1)}_{10}(k_x,q) 
&=& \frac{M^{(a)}}{\varepsilon_{v}^{(a)}} + \frac{M^{(b)}}{\varepsilon_{v}^{(b)}} 
\nonumber\\
&=& \frac{gn}{2} \frac{2\alpha}{\pi}
\bigg[- \frac{4E_r\beta_{10}(q)\zeta_{1010}(k_x,q)}
{\hbar\omega-\Delta_{10}(q)} 
\nonumber\\&& +\,                 
\frac{2E_r\beta_{10}(q+k_x)\zeta_{0000}(k_x,q)}
{\hbar\omega-\Delta_{10}(q+k_x)}
\bigg].
\ees
where $n=N/L^3$. Moreover, one has
\be
C_{01}^{(1)}(k_x,q) = C_{10}^{(1)}(-k_x,q).
\ee

\begin{figure}[t]
	\includegraphics[width=0.7\linewidth]{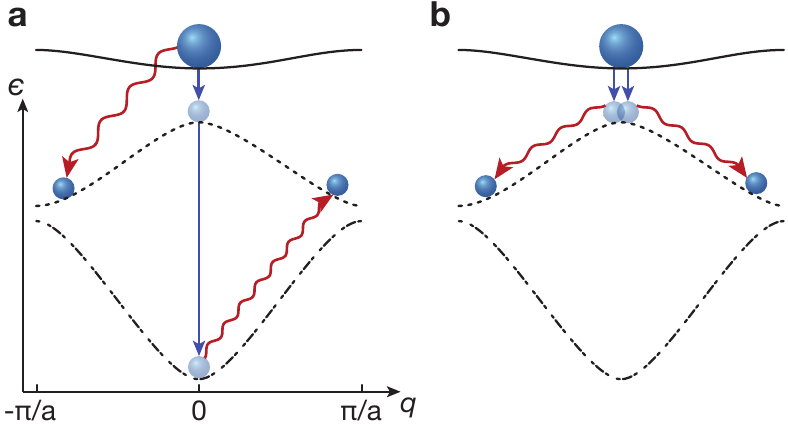}
	\includegraphics[width=0.7\linewidth]{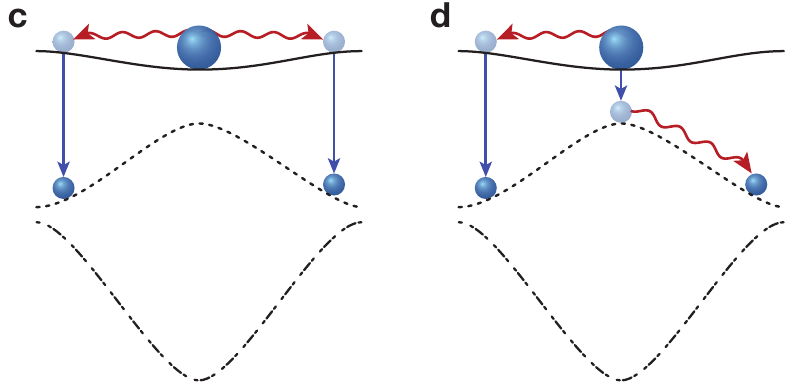}
	\centering
	\caption{\label{fig:C11} Dominant single-photon second-order processes
		contributing to the matrix element $C^{(2)}_{11}(\bk_x)$. Like Fig.~\ref{fig:C01}.} 
\end{figure}

\begin{figure}[t]
	\includegraphics[width=0.7\linewidth]{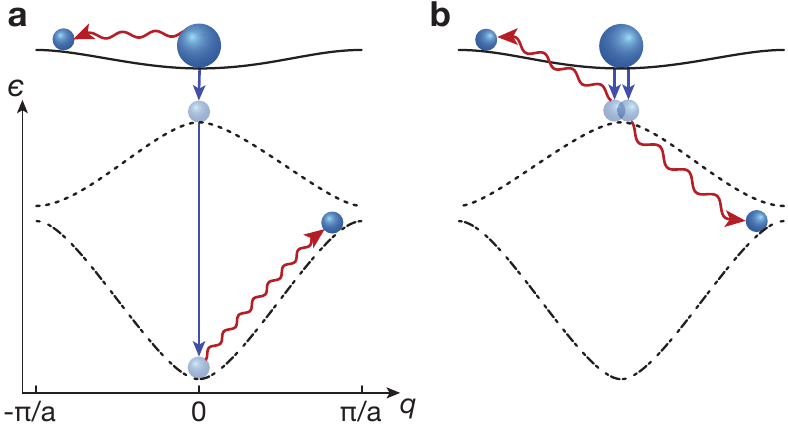}
	\includegraphics[width=0.7\linewidth]{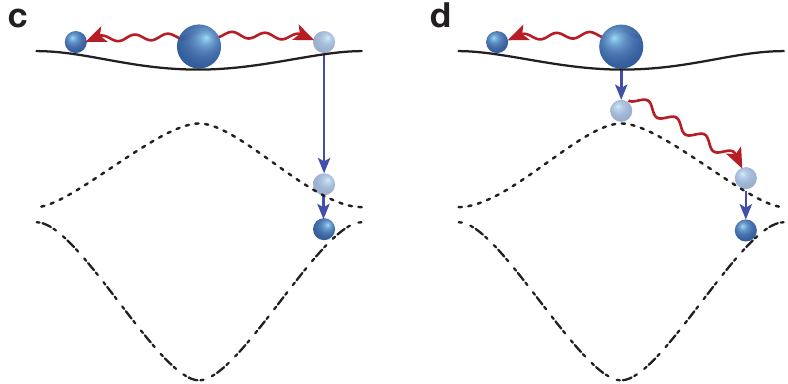}
	\centering
	\caption{\label{fig:C20} Dominant single-photon second-order processes
		contributing to the matrix element $C^{(2)}_{20}(\bk_x)$. Like Fig.~\ref{fig:C01}.} 
\end{figure}

When computing the effective single-photon scattering matrix element
(\ref{eq:C1ab}), the contributions from process (a) and (b) have opposite sign.
Accidentally both contributions have almost the same absolute value, so that
the whole matrix element is much smaller than the ones for the individual 
processes (a) and (b). As a result, subdominant processes that involve virtual 
intermediate states of large energy separation [i.e.\ that violate the 
selection principle (iii)] become relevant. There is one such process, 
which is depicted in Fig.~\ref{fig:C01}(c). It can be computed in a similar fashion
as the processes (a) and (b) and gives rise to a significant correction

to $C^{(1)}_{10}(k_x,q)$, which reads
\be
\frac{gn}{2} \frac{2\alpha}{\pi}\frac{2E_r\beta_{01}(q-k_x)\zeta_{1100}(k_x,q)}
{\hbar\omega+\Delta_{10}(q-k_x)}.
\ee

When computing the numerical values for the resulting scattering rates, we will 
see that the third term matters indeed. There is also another subleading term 
appearing in first order contributing to $C^{(1)}_{10}(k_x,q)$, the single-photon 
scattering process described by the matrix element
$B_{1\bq+\bk,0\bq-\bk,0\bq,0\bq}^{(-1)} N $. This matrix element is rather small, 
since $\zeta_{1000}(k_x,q)$ is nonzero only due to interaction processes that do 
not happen on-site. We have verified that its impact is negligible.

In the next order of perturbation theory, the processes sketched in 
Fig.~\ref{fig:C11} contribute to the matrix elements $C_{11}(k_x)$, whereas
those depicted in Fig.~\ref{fig:C20} contribute to the matrix elements
$C_{20}(k_x)$. We have computed these two-photon matrix elements, which are 
proportional to $\alpha^2$, and took them into account in Fig.~2(e) of the main 
text. 

\clearpage
\end{document}